\documentclass[twocolumn,reprint,amsmath,amssymb,aps,superscriptaddress,pre]{revtex4-1}

\usepackage[dvipsnames]{xcolor}
\usepackage{graphicx}
\usepackage{hyperref}
\hypersetup{colorlinks=true,allcolors=NavyBlue}

\DeclareMathOperator{\sign}{sign}
\DeclareMathOperator{\erf}{erf}

\DeclareMathOperator{\relu}{ReLU}
\DeclareMathOperator{\He}{He}
\let\var\relax
\DeclareMathOperator{\var}{var}
\DeclareMathOperator{\cov}{cov}
\DeclareMathOperator*{\extr}{extr}
\newcommand{\T}{\mathrm{T}}

\begin{document}

\title{Activation function dependence of the storage capacity of treelike neural networks}

\author{Jacob A. Zavatone-Veth}
\email{jzavatoneveth@g.harvard.edu}
\affiliation{Department of Physics, Harvard University, Cambridge, Massachusetts 02138, USA}

\author{Cengiz Pehlevan}
\email{cpehlevan@seas.harvard.edu}
\affiliation{John A. Paulson School of Engineering and Applied Sciences, Harvard University, Cambridge, Massachusetts 02138, USA}
\affiliation{Center for Brain Science, Harvard University, Cambridge, Massachusetts 02138, USA}

\date{\today}

\begin{abstract}
The expressive power of artificial neural networks crucially depends on the nonlinearity of their activation functions. Though a wide variety of nonlinear activation functions have been proposed for use in artificial neural networks, a detailed understanding of their role in determining the expressive power of a network has not emerged. Here, we study how activation functions affect the storage capacity of treelike two-layer networks. We relate the boundedness or divergence of the capacity in the infinite-width limit to the smoothness of the activation function, elucidating the relationship between previously studied special cases. Our results show that nonlinearity can both increase capacity and decrease the robustness of classification, and provide simple estimates for the capacity of networks with several commonly used activation functions. Furthermore, they generate a hypothesis for the functional benefit of dendritic spikes in branched neurons. 
\end{abstract}

\maketitle

The expressive power of artificial neural networks is well-known \cite{cybenko1989approximation,hornik1989multilayer,zhang2016understanding,poole2016exponential}, but a complete theoretical account of how their remarkable abilities arise is lacking \cite{lecun2015deep,goodfellow2016deep,belkin2019reconciling,zdeborova2020understanding}. In particular, though a diverse array of nonlinear activation functions have been employed in neural networks \cite{lecun2015deep,goodfellow2016deep,panigrahi2019effect,ramachandran2017searching,baldassi2019properties,barkai1992broken,engel1992storage,engel2001statistical}, our understanding of the relationship between activation function choice and computational capability is incomplete \cite{panigrahi2019effect,ramachandran2017searching,baldassi2019properties,mozeika2020space}. Methods from the statistical mechanics of disordered systems have enabled the interrogation of this link in several special cases \cite{gardner1988space,gardner1988optimal,engel1992storage,barkai1992broken,baldassi2019properties,talagrand2003spin,engel2001statistical,monasson1995weight,mozeika2020space}, but these previous works have not yielded a general theory.

In this Letter, we characterize how pattern storage capacity depends on activation function in a tractable two-layer network model known as the treelike committee machine (henceforth TCM). In addition to their uses in machine learning, TCMs have been used to model nonlinear computations in dendrite-bearing neurons \cite{poirazi2003pyramidal,poirazi2020illuminating}. We find that the storage capacity of a TCM remains finite in the infinite-width limit provided that the activation function is weakly differentiable, and it and its weak derivative are square-integrable with respect to Gaussian measure. For example, the capacity with sign activation functions diverges, while that with rectified linear unit or error function activations is finite. We predict that nonlinearity should increase capacity, but may reduce the robustness of classification. These connections between expressive power and smoothness begin to shed light on the influence of activation functions on the capabilities of neural networks and branched neurons.

\begin{figure}[b]
    \centering
    \includegraphics[width=\columnwidth]{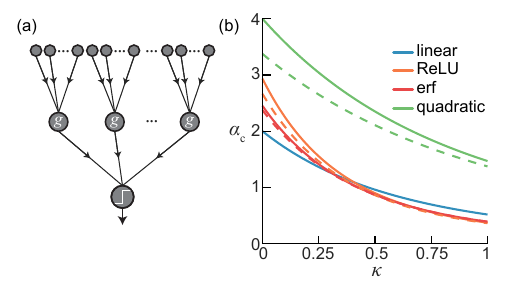}
    \caption{Pattern storage in treelike committee machines. (a) Network architecture. (b) Capacity $\alpha_{c}$ as a function of margin $\kappa$ for several common activation functions. Solid and dashed lines indicate estimates of the capacity under replica-symmetric and one-step replica-symmetry-breaking ans{\"a}tze, respectively. }
    \label{fig:diagram}
\end{figure}

\emph{The treelike committee machine}---The TCM is a two-layer neural network with $N$ inputs divided among $K$ hidden units into disjoint groups of $N/K$ and binary outputs (Figure \ref{fig:diagram}a) \cite{baldassi2019properties,engel1992storage,barkai1992broken,engel2001statistical,monasson1995weight}. For a hidden unit activation function $g$, a set of hidden unit weight vectors $\{\mathbf{w}_{j} \in \mathbb{R}^{N/K}\}_{j=1}^{K}$, a readout weight vector $\mathbf{v} \in \mathbb{R}^{K}$, and a threshold $\vartheta \in \mathbb{R}$, its output is given as
\begin{align}
    y(\mathbf{x}) &= \sign(s(\mathbf{x})) \quad \mathrm{for}
    \\
    s(\mathbf{x}; \{\mathbf{w}_{j}\}, \mathbf{v}, \vartheta) &= \frac{1}{\sqrt{K}} \sum_{j=1}^{K} v_{j} g\left(\frac{\mathbf{w}_{j} \cdot \mathbf{x}_{j}}{\sqrt{N/K}}\right) - \vartheta,
\end{align}
where $\mathbf{x}_{j}$ denotes the vector of inputs to the $j^{th}$ hidden unit. In this model, the readout weight vector and threshold are fixed, and only the hidden unit weights are learned. The perceptron can thus be viewed as the special case of a TCM with identity activation functions and equal readout weights \cite{gardner1988space,gardner1988optimal}. 

\emph{Statistical mechanics of pattern storage}---To characterize this network's ability to classify a random dataset of $P$ examples subject to constraints on the hidden unit weights imposed by a probability measure $\rho$, we define the Gardner volume \cite{gardner1988space,gardner1988optimal}
\begin{align}
    Z = \int d\rho(\{\mathbf{w}_{j}\}) \prod_{\mu=1}^{P} \Theta\left( y^{\mu} s(\mathbf{x}^{\mu}; \{\mathbf{w}_{j}\}, \mathbf{v}, \vartheta) - \kappa \right), 
\end{align}
which measures the fractional volume in weight space such that all examples are classified correctly with margin at least $\kappa$. We consider ``spherical" committee machines, in which the hidden unit weight vectors lie on the sphere of radius $(N/K)^{1/2}$ \cite{gardner1988space,engel1992storage,barkai1992broken,baldassi2019properties,monasson1995weight,engel2001statistical,gardner1988optimal,talagrand2003spin}. As in most studies of the Gardner volume, we consider a dataset in which the components of the inputs and the target outputs are independent and identically distributed as $x_{jk}^{\mu} = \pm 1$ and $y^{\mu} = \pm 1$ with equal probability \cite{gardner1988space,gardner1988optimal,engel1992storage,talagrand2003spin,engel2001statistical,barkai1992broken,baldassi2019properties,monasson1995weight}.

We will study a sequential infinite-width limit in which we first take $N,P\to\infty$ with load $\alpha \equiv P/N = \mathcal{O}(1)$ and then take $K \to \infty$ (see Appendix \ref{app:sec:finite}). The infinite-width limit is of both theoretical and practical interest, as extremely wide networks are now commonly used in applications \cite{belkin2019reconciling,jacot2018neural,bordelon2020spectrum,panigrahi2019effect}. In this limit, we expect the free entropy per weight $f = N^{-1} \log Z$ to be self-averaging, and for there to exist a critical load $\alpha_{c}$, termed the capacity, below which the classification task is solvable with probability one and above which $Z$ vanishes \cite{gardner1988space,gardner1988optimal,talagrand2003spin,engel2001statistical}. The special case of this model with sign activation functions was intensively studied in the late 20\textsuperscript{th} century, showing that the capacity diverges as $K \to \infty$ \cite{engel1992storage,barkai1992broken,monasson1995weight,mitchison1989bounds} \footnote{This divergence is slow, with $\alpha_{c} \sim \sqrt{\log K}$ \cite{monasson1995weight,mitchison1989bounds}; we provide a detailed discussion of this and other finite-size effects in Appendix \ref{app:sec:finite}.}. In contrast, \citet{baldassi2019properties} showed in a recent Letter that the capacity with rectified linear unit (ReLU) activations remains bounded in the infinite-width limit. Our primary objective in this work is to identify the class of activation functions for which the capacity remains finite.

We begin our analysis by specifying our choice of general constraints on the activation function, readout weights, and threshold. We will require the $K \to \infty$ limit to be well-defined in the sense that the output preactivation $s$ has finite variance. In this limit, the central limit theorem implies that the hidden unit preactivations converge in distribution to a collection of independent Gaussian random variables \cite{pollard2002user}. Therefore, the activation function $g$ must lie in the Lebesgue space $\mathcal{L}^{2}(\gamma)$ of functions that are square-integrable with respect to the Gaussian measure $\gamma$ on the reals. Furthermore, as $\var(s) \propto \Vert \mathbf{v} \Vert_{2}^{2}/K$, we must have $\Vert \mathbf{v} \Vert_{2} = \mathcal{O}(\sqrt{K})$. As $\Vert \mathbf{v} \Vert_{2}$ sets the effective scale of $\vartheta$ and $\kappa$ but does not affect the zero-margin capacity, we fix $\Vert \mathbf{v} \Vert_{2} = \sqrt{K}$. To ensure that $s$ has mean zero, we set $\vartheta = K^{-1/2} (\mathbb{E} g) \sum_{j=1}^{K} v_{j}$, where $\mathbb{E}g = \int d\gamma\, g$ is the average hidden unit activation. This choice maximizes the capacity for the symmetric datasets of interest (see Appendices \ref{app:sec:gardner}, \ref{app:sec:rs}, and \ref{app:sec:1rsb}), and generalizes the conditions on $\mathbf{v}$ and $\vartheta$ considered in previous works \cite{baldassi2019properties,engel1992storage,barkai1992broken,monasson1995weight}.

To compute the limiting quenched free entropy, we apply the replica trick, which exploits a limit identity for logarithmic averages and a non-rigorous interchange of limits to write
\begin{align}
    f = \lim_{n \downarrow 0} \lim_{K \to \infty} \lim_{N \to \infty} \frac{1}{nN} \log \mathbb{E}_{\mathbf{x},y} Z_{N,\alpha N,K}^{n},
\end{align}
where the validity of analytic continuation of the moments from positive integer $n$ to $n \downarrow 0$ is assumed \cite{gardner1988space,talagrand2003spin,mezard1987spin}. This calculation is standard, and we defer the details to Appendix \ref{app:sec:gardner}

In this limit, the quenched free entropy can be expressed using the method of steepest descent as an extremization over the Edwards-Anderson order parameters $q_{j}^{ab} = (K/N) \mathbf{w}_{j}^{a} \cdot \mathbf{w}_{j}^{b}$ \cite{gardner1988space,talagrand2003spin,mezard1987spin}, which represent the average overlap between the preactivations of the $j^{th}$ hidden unit in two different replicas $a$ and $b$. Under a replica- and hidden-unit-symmetric (RS) ansatz $q_{j}^{ab} = q$, one finds that
\begin{align}
    f_{\textrm{RS}} = \extr_{q} \bigg\{ & \alpha \int d\gamma(z)\, \log H\left(\frac{\kappa + \sqrt{\tilde{q}(q)} z}{\sqrt{\sigma^2 - \tilde{q}(q)}}\right) \nonumber \\&+ \frac{1}{2} \left[\frac{q}{1-q} + \log(1-q) \right] \bigg\},
\end{align}
where $H(z) = \int_{z}^{\infty} d\gamma(x)$ is the Gaussian tail distribution function, $\sigma^2 = \mathbb{E} g^2 - (\mathbb{E} g)^2$ is the variance of the activation, and 
\begin{align} \label{eqn:qtilde}
    \tilde{q}(q) = \cov\left[g(x),g(y) \,:\, \begin{bmatrix} x \\ y \end{bmatrix} \sim \mathcal{N}\left(0, \begin{bmatrix} 1 & q \\ q & 1 \end{bmatrix} \right) \right]
\end{align}
is an effective order parameter describing the average overlap between the activations of a given hidden unit in two different replicas. This expression for $f_{\textrm{RS}}$ is equivalent to that given in \cite{baldassi2019properties} for ReLU activations, but we adopt a different definition for the effective order parameter that has a clearer statistical interpretation.

To find the replica-symmetric capacity $\alpha_{\textrm{RS}}$, one must take the limit $q \uparrow 1$ in the saddle point equation that defines the extremum with respect to $q$, as the Gardner volume tends to zero in this limit \cite{gardner1988space,gardner1988optimal,engel1992storage,engel2001statistical,baldassi2019properties,barkai1992broken}. As $q \uparrow 1$, $\tilde{q} \uparrow \sigma^2$, but the asymptotic properties of $\tilde{q}$ as a function of $\varepsilon \equiv 1 - q$ depend on the choice of activation function. Making the general ansatz that $\sigma^2 - \tilde{q} \sim \varepsilon^{\ell}$ for some $\ell > 0$, we find that $\alpha_{\textrm{RS}} \sim \varepsilon^{\ell-1}$ (see Appendix \ref{app:sec:rs}). Therefore, the RS capacity diverges if $\ell < 1$ and vanishes if $\ell > 1$, while the boundary case $\ell = 1$ is special in that the capacity is bounded but non-vanishing. For the special cases of $\sign(x)$ and $g(x) = \relu(x)$, this behavior was noted by \citet{baldassi2019properties}. For sign, one has $\sigma^2 - \tilde{q} \sim \sqrt{\varepsilon}$, and $\alpha_{\textrm{RS}}$ diverges in the infinite-width limit, while for ReLU, $\sigma^2 - \tilde{q} \sim \varepsilon$, and $\alpha_{\textrm{RS}}$ remains finite. However, \cite{baldassi2019properties} and other previous studies \cite{engel1992storage,barkai1992broken} relied on direct computation of the effective order parameters for all values of $q$, which is not tractable for most activation functions, and does not yield general insight.

\emph{Asymptotics of the effective order parameter}---To understand the asymptotic behavior of $\tilde{q}(q)$ as $q \uparrow 1$ for general activation functions $g$, we apply tools from the theory of Gaussian measures \cite{bogachev1998gaussian}. As $g$ is in $\mathcal{L}^{2}(\gamma)$ by assumption, it has a Fourier-Hermite series $g(x) = \sum_{k=0}^{\infty} g_{k} \He_{k}(x)$,
where $\{\He_{k}\}$ is the set of orthonormal Hermite polynomials (see Appendix \ref{app:sec:gauss}). We note that the $\mathcal{L}^{2}(\gamma)$ norm of $g$ can then be written as $\Vert g \Vert_{\gamma}^{2} = \sum_{k=0}^{\infty}g_k^2$, and that $g_{0} = \mathbb{E} g$. To express $\tilde{q}(q)$ in terms of these coefficients, we recall the Mehler expansion of the standard bivariate Gaussian density $\varphi(x,y;q)$ \cite{kibble1945extension,tong2012multivariate}: $\varphi(x,y;q) = \varphi(x) \varphi(y) \sum_{k=0}^{\infty} q^{k} \He_{k}(x) \He_{k}(y)$,
where $\varphi(x) = \exp(-x^2/2)/\sqrt{2\pi}$ is the univariate Gaussian density. Then, we can evaluate the expectation in (\ref{eqn:qtilde}), yielding $\tilde{q}(q) + g_{0}^{2} = \sum_{k=0}^{\infty} g_{k}^{2} q^{k}$,
%
%
which, by Abel's theorem, is a bounded, continuous function of $q \in (-1,1]$ because $\tilde{q}(1) + g_{0}^{2} = \Vert g \Vert_{\gamma}^{2}$ is finite. Writing $q \equiv 1-\varepsilon$, we expand $(1-\varepsilon)^{k}$ in a binomial series and formally interchange the order of summation to obtain $\tilde{q}(\varepsilon) + g_{0}^{2} = \sum_{l=0}^{\infty} \frac{(-\varepsilon)^{l}}{l!} \sum_{k=l}^{\infty} (k)_{l} g_{k}^{2}$,
where $(k)_{l} = k (k-1) \cdots (k-l+1)$ is the falling factorial. We recognize the sums over $k$ as the norms of the weak derivatives of $g$, which have formal Fourier-Hermite series $g^{(l)}(x) = \sum_{k=l}^{\infty} g_{k} \sqrt{(k)_{l}} \He_{k-l}(x)$,
which follow from the recurrence relation $\He_{k}'(x) = \sqrt{k} \He_{k-1}(x)$ \cite{bogachev1998gaussian}. Therefore, $\tilde{q}$ admits a formal power series expansion in $\varepsilon$ as
\begin{align}
    \tilde{q}(\varepsilon) + g_{0}^{2} = \sum_{l=0}^{\infty} \frac{(-1)^{l}}{l!} \Vert g^{(l)} \Vert_{\gamma}^{2}\, \varepsilon^{l}. 
\end{align}

For the RS capacity to remain bounded, we merely require that the first two terms in this series are finite, not for the series to converge at any higher order for non-vanishing $\varepsilon$. Therefore, the RS capacity is finite for once weakly-differentiable activations $g$ such that the $\mathcal{L}^{2}$ norms of the function and its weak derivative with respect to Gaussian measure,  $\Vert g \Vert_{\gamma}$ and $\Vert g' \Vert_{\gamma}$, are finite. This class of functions is precisely the Sobolev class $\mathcal{H}^{1}(\gamma)$ \cite{bogachev1998gaussian}. We provide additional background material on $\mathcal{H}^{1}(\gamma)$ and weak differentiability in Appendix \ref{app:sec:gauss}.

\emph{Storage capacity}---For any activation function in the class $\mathcal{H}^{1}(\gamma)$, we find that
\begin{align}\label{eqn:rscapacity}
    \alpha_{\textrm{RS}}(\kappa) = \frac{\Vert g' \Vert_{\gamma}^{2}}{\sigma^{2}} \alpha_{\textrm{G}}\left(\frac{\kappa}{\sigma}\right), 
\end{align}
where
\begin{align}
    \alpha_{\textrm{G}}(\kappa) = \left[\int_{-\kappa}^{\infty} d\gamma(z)\, (\kappa + z)^2\right]^{-1}
\end{align}
is Gardner's formula for the perceptron capacity \cite{gardner1988space} (see Appendix \ref{app:sec:rs}). In terms of Fourier-Hermite coefficients, we have $\sigma^2 = \sum_{k=1}^{\infty} g_{k}^{2}$ and $\Vert g' \Vert_{\gamma}^{2} = \sum_{k=1}^{\infty} k g_{k}^{2}$. Thus, we have $\Vert g' \Vert_{\gamma}^{2} \geq \sigma^2$, with equality if and only if all nonlinear terms (those corresponding to Hermite polynomials of degree two or greater) vanish. Therefore, introducing nonlinearity always increases the zero-margin RS capacity. However, as $\alpha_{\textrm{G}}(\kappa)$ is a monotonically decreasing function, the capacity at large margins can be reduced by nonlinearity if $\sigma < 1$. We note that the zero-margin capacity is invariant under rescaling of the activation function and hidden unit weights as $g \mapsto c_{1} g$, $\mathbf{v} \mapsto c_{2} \mathbf{v}$ for some constants $c_{1}$ and $c_{2}$. For finite margin, rescaling can increase or decrease the capacity by changing $\sigma$. Thus, in the sense of classification margin, introducing nonlinearity or re-scaling can reduce the robustness of classification.

Using this result, we can characterize the RS capacity of wide TCMs for several commonly-used activation functions (see Appendix \ref{app:sec:examples}). For a linear activation function, our result reduces to Gardner's perceptron capacity \cite{gardner1988space}, which is expected given the equivalence between such a TCM and the perceptron in the $K\to\infty$ limit. As the sign function is not weakly differentiable, we recover the result that the capacity diverges \cite{engel1992storage,barkai1992broken,monasson1995weight}. ReLU is weakly differentiable, and we recover the result of \cite{baldassi2019properties} that $\alpha_{\textrm{RS}} = 2\pi/(\pi-1) \simeq 2.93388$. Considering sigmoidal activations, we find that $\alpha_{\textrm{RS}} = 2\arcsin(2/3)/\pi \simeq 2.45140$ for the error function, while $\alpha_{\textrm{RS}} \simeq 2.35561$ for the hyperbolic tangent and the logistic. As an example of a non-monotonic activation function, we consider a quadratic, which yields $\alpha_{\textrm{RS}} = 4$. We plot the RS capacity as a function of margin for these activation functions in Figure \ref{fig:diagram}b, illustrating how nonlinearity can reduce the large-margin capacity while increasing the zero-margin capacity.

However, for nonlinear activation functions, one generically expects the energy landscape to become locally non-convex, and for replica symmetry breaking (RSB) to occur \cite{mezard1987spin,talagrand2003spin,engel1992storage,barkai1992broken,baldassi2019properties,engel2001statistical}. The RS estimate of the capacity is therefore only an upper bound, and one must account for RSB effects in order to obtain a more accurate estimate \cite{mezard1987spin,talagrand2003spin,engel1992storage,barkai1992broken,baldassi2019properties,engel2001statistical,monasson1995weight}. To that end, we have calculated the capacity under a one-step replica-symmetry-breaking (1-RSB) ansatz, extending the results of earlier work \cite{engel1992storage,barkai1992broken,baldassi2019properties} to arbitrary activation functions. Under the 1-RSB ansatz, the replicas are divided into groups of size $m$, with inter-group overlap $q_{0}$ and intra-group overlap $q_{1}$. Then, the capacity is extracted by taking the limit $q_{1} \uparrow 1$, $m \downarrow 0$, with $r \equiv m / (1-q_{1})$ finite \cite{engel1992storage,barkai1992broken,baldassi2019properties,engel2001statistical,mezard1987spin}.

As detailed in Appendix \ref{app:sec:1rsb}, this calculation yields an expression for $\alpha_{\textrm{1-RSB}}$ as the solution to a two-dimensional minimization problem over $q_{0}$ and $r$. Importantly, the finite-capacity condition at 1-RSB is the same as that with RS. For functions in $\mathcal{H}^{1}(\gamma)$, the resulting minimization problem must usually be solved numerically, hence we give results for only a few tractable examples. RSB does not occur for linear activation functions \cite{talagrand2003spin,gardner1988space,gardner1988optimal,shcherbina2003rigorous}. For ReLU, we obtain $\alpha_{\textrm{1-RSB}} \simeq 2.66428$ at $(q_{0}^{\ast},r^{\ast}) \simeq (0.75716, 16.6374)$, which is consistent with the result of \citet{baldassi2019properties} (see \footnote{In the published version of their Letter, \citet{baldassi2019properties} reported a value of $\alpha_{\textrm{1-RSB}} \simeq 2.92$. After the appearance of our work in preprint form, they found that this result was incorrect (see Appendix \ref{app:sec:examples}); their revised estimate of $\alpha_{\textrm{1-RSB}} \simeq 2.6643$ agrees with our results.}). For erf, we obtain $\alpha_{\textrm{1-RSB}} \simeq 2.37500$ at $(q_{0}^{\ast},r^{\ast}) \simeq (0.75463, 7.75682)$. Finally, for the quadratic, we have $\alpha_{\textrm{1-RSB}} \simeq 3.37466$ at $(q_{0}^{\ast},r^{\ast}) \simeq (0.28452, 6.39299)$. In Figure \ref{fig:diagram}, we plot the 1-RSB capacity for these activation functions at nonzero margins. The gap between the RS and 1-RSB results for the quadratic is larger than that for erf or ReLU, both in the numerical value of the capacity and in the difference between $q_{0}^{\ast}$ and $q_{1}^{\ast}$. Though the capacities at 1-RSB are reduced relative to the RS result, their ordering for these activation functions is preserved.

\begin{figure}[tb]
    \centering
    \includegraphics[width=\columnwidth]{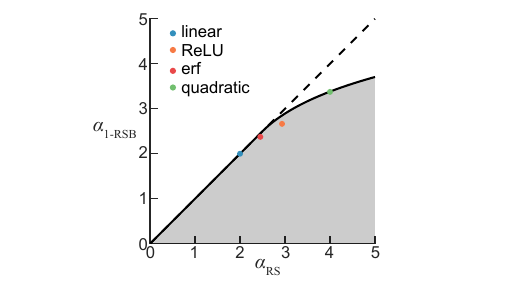}
    \caption{The accessible region in ($\alpha_{\textrm{RS}}$, $\alpha_{\textrm{1-RSB}}$)-space defined by the $q_{0}=0$ bound. The allowed region is shaded in gray, and the locations of the four example activation functions for which we estimate $\alpha_{\textrm{1-RSB}}$ are indicated by colored dots.}
    \label{fig:1rsb_space}
\end{figure}

For general activation functions in $\mathcal{H}^{1}(\gamma)$, we can obtain informative upper bounds on $\alpha_{\textrm{1-RSB}}$ by considering candidate solutions with fixed values of the inter-block overlap $q_{0}$. From $q_{0}\uparrow 1$, we have $\alpha_{\textrm{1-RSB}} \leq \alpha_{\textrm{RS}}$. As shown in Appendix \ref{app:sec:1rsb}, we can also obtain an upper bound for $\alpha_{\textrm{1-RSB}}$ at zero margin as a function of $\alpha_{\textrm{RS}}$ by taking $q_{0} = 0$ and optimizing over $r$ alone. For $\alpha_{\textrm{RS}} \leq 5/2$, these two bounds coincide, while the $q_{0}=0$ bound is tighter for $\alpha_{\textrm{RS}} > 5/2$. In particular, for $\alpha_{\textrm{RS}} \gg 1$, this yields $\alpha_{\textrm{1-RSB}} = \mathcal{O}(\log \alpha_{\textrm{RS}})$. The $q_{0}=0$ bound allows us to define an accessible region in ($\alpha_{\textrm{RS}}$, $\alpha_{\textrm{1-RSB}}$)-space, as illustrated in Figure \ref{fig:1rsb_space}. Our numerical estimates for the 1-RSB capacities of ReLU, erf, and the quadratic all lie within this allowed area, and are relatively close to the $q_{0}=0$ bound (see Appendix \ref{app:sec:examples}).

These bounds suggest that RSB strongly affects the capacity for activation functions with large derivative norm and thus large $\alpha_{\textrm{RS}}$. This is illustrated by the example of Hermite polynomial activation functions. For $g(x) = \He_{k}(x)$, we have $\alpha_{\textrm{RS}}(\kappa=0) = 2k$, hence one can obtain an arbitrarily large, but finite, zero-margin RS capacity by taking $k \gg 1$. However, as shown in Appendix \ref{app:sec:examples}, the 1-RSB capacity grows extremely slowly---sub-logarithmically---with degree. This result is sensible given the oscillatory nature of high-degree Hermite polynomials, which one expects to yield a highly non-convex energy landscape.

\emph{Discussion}---We have shown that the storage capacity of treelike committee machines with activation functions in $\mathcal{H}^{1}(\gamma)$ remains bounded in the infinite-width limit. Our results follow from a replica analysis of the Gardner volume, with the capacity given by a simple closed-form expression under a replica-symmetric ansatz and a two-dimensional minimization problem with one-step replica-symmetry-breaking. Depending on the activation function, a fully accurate determination of the capacity would likely require higher levels in the Parisi hierarchy of replica-symmetry-breaking ans{\"a}tze \cite{mezard1987spin}. Furthermore, it can be challenging to rigorously prove that the capacity results obtained using the replica method at any level of the Parisi hierarchy are correct \cite{talagrand2003spin,mezard1987spin,shcherbina2003rigorous,ding2019capacity,aubin2019committee}. With these caveats in mind, our results begin to elucidate how nonlinear activation functions affect the ability of neural networks to robustly solve classification problems.

Though our analysis focused on a regime in which the input distribution is symmetric, inputs in both biological and artificial neural networks are often only sparsely active \cite{knoblauch2010memory,willmore2001characterizing}. Our analysis of the RS capacity can be extended to this regime (see Appendix \ref{app:sec:sparse}), following \citet{gardner1988space}'s work on the perceptron. Provided that the input and target output distributions are not both infinitely sparse, the condition for the capacity to remain finite in the infinite-width limit remains the same. However, if the activation function can be linearized about zero, the zero-margin capacity for a symmetric target distribution decreases to that of the perceptron in the limit of very sparsely active inputs. This holds, for instance, for erf or tanh, but not for ReLU, for which the zero-margin capacity is independent of sparsity. This example illustrates how introducing simple yet realistic forms of data structure can affect pattern storage. Investigating how other forms of data structure affect storage capacity will be an important objective for future work \cite{shinzato2008perceptron,pastore2020structured,goldt2020modelling,zdeborova2020understanding}.

In addition to its use as a model system in machine learning, the TCM has been proposed as an abstract model for computation in dendrite-bearing neurons \cite{poirazi2003pyramidal,poirazi2020illuminating,breuer2014statistical}. In this application, each hidden unit represents a dendritic unit that integrates some set of synaptic inputs to generate a signal that is transmitted to the soma, which in turn generates a ``spike" if the total current exceeds a threshold \cite{poirazi2003pyramidal,poirazi2020illuminating}. The most striking form of nonlinearity observed in measurements of dendritic signal processing is the generation of dendritic spikes \cite{gidon2020dendritic,payeur2019classes}. Though it is difficult to argue that biological nonlinearities can be infinitely sharp, previous works have modeled dendritic spikes using non-weakly-differentiable activation functions \cite{poirazi2003pyramidal,poirazi2020illuminating,breuer2014statistical}. Our work therefore generates a hypothesis for the functional benefit of dendritic spikes: non-smooth dendritic nonlinearities allow the capacity to grow without bound as the number of branches increases and to remain robustly large even when inputs are very sparse. It will be interesting to test this hypothesis using computational models that incorporate greater biophysical realism \cite{poirazi2020illuminating}.

The Gardner volume is agnostic to the choice of learning algorithm used to train the weights of the network. This feature makes it a general approach to studying storage capacity, but means that it can provide only limited insight into the practical realizability of the extant solutions \cite{engel1992storage,engel2001statistical,barkai1992broken,baldassi2019properties,malach2019deeper}. As a result, it is challenging to directly test theories of the Gardner volume. It is nevertheless possible to experimentally falsify such theories; we have failed to do so (see Appendix \ref{app:sec:examples}). More broadly, this distinction between satisfiability and learnability, combined with its dependence on data and focus on perfect classification, means that the Gardner volume is one of many metrics that should be considered in evaluating activation function choice \cite{knoblauch2010memory,panigrahi2019effect,malach2019deeper,ramachandran2017searching}. 
In a recent study of least-squares function approximation by wide fully-connected networks, \citet{panigrahi2019effect} have shown that the speed and robustness of gradient descent learning is related to activation function smoothness. Their result is suggestively similar to that of this Letter, though it is as yet unclear whether a similar link between smoothness and trainability exists for treelike networks.

In this work, we have studied the activation-function-dependence of the storage capacity of wide TCMs. This network architecture is particularly convenient to study in the infinite-width limit, but it is far removed from the deep networks used in practical applications \cite{lecun2015deep}. As a more realistic model, one could consider a fully-connected committee machine (FCM), in which each hidden unit is connected to the full set of inputs. Prior work on such networks with sign activation functions suggests that some qualitative aspects of the behavior of TCMs should still hold true \cite{engel1992storage,barkai1992broken,urbanczik1997storage}. However, FCMs possess a symmetry with respect to permutation of the hidden units, which is broken at loads below the RS capacity \cite{barkai1992broken}. This phenomenon and the presence of correlations between hidden units complicate the study of their infinite-width limit. Accurate determination of how FCM storage capacity depends on activation function will therefore require further work, in which the insights developed in this study should prove broadly useful.

\emph{Acknowledgements}---J. A. Z.-V. acknowledges support from the NSF-Simons Center for Mathematical and Statistical Analysis of Biology at Harvard and the Harvard Quantitative Biology Initiative. C. P. thanks the Harvard Data Science Initiative, Google, and Intel for support. 

\bibliography{committee_refs}

\newpage 

\onecolumngrid

\appendix

\renewcommand{\thefigure}{S\arabic{figure}}
\setcounter{figure}{0}

\section{Finite-size effects in treelike committee machines}\label{app:sec:finite}

The treelike committee machine has three relevant scales: the total number of inputs $N$, the number of hidden units $K$, and the number of inputs per hidden unit $N/K$. We note that we implicitly assume that $N \geq K$ throughout, and ignore whether these three scales are truly integer valued. In our calculations, we consider a limit in which $N$ is first taken to infinity for fixed $K$ (hence $N/K$ tends to infinity), and then $K$ is taken to infinity. In this appendix, we discuss how this limit relates to alternative infinite-size limits, and how finite size effects in each of these scales might affect our results. 

For finite $N$, a phase transition in the Gardner volume is not possible. Instead, the fraction of realizable dichotomies decays smoothly from one to zero with increasing load. A rigorous analysis of this effect for the perceptron with zero margin ($\kappa=0$) is provided by Cover's theorem \cite{cover1965geometrical}. Cover's theorem identifies the critical load $\alpha_{c}$ for finite $N$ as that for which half of all dichotomies are realizable, and shows that this value is independent of $N$. As $N\to\infty$, the decay at $\alpha_{c}$ becomes infinitely sharp. The combinatorial methods used by Cover are not easily extensible to multilayer networks \cite{cover1965geometrical,mitchison1989bounds,pastore2020structured}, rendering rigorous analysis of finite-$N$ effects more difficult. However, one expects these finite-size effects to be qualitatively similar even in two-layer models \cite{engel1992storage,engel2001statistical,barkai1992broken,mitchison1989bounds}. 

Taking the limit $K\to\infty$ substantially simplified our calculations, as the resulting limiting distribution of the output preactivation is Gaussian. One could systematically study finite-$K$ corrections to this limit by expanding the distribution of the output preactivation in powers of $K^{-1/2}$ as an Edgeworth series \cite{kolassa2006series}. However, studying the behavior of networks with small $K$ is analytically challenging for general choices of activation function, as one must deal directly with the full distributions of the hidden unit activations rather than just their first few cumulants. 

For sign activation functions, one can derive a combinatorial expression for the finite-$K$ capacity thanks to the fact that the required distributions are discrete \cite{engel1992storage,barkai1992broken,monasson1995weight}. In particular, the capacity is a monotonically increasing function of $K$, and scales with $\sqrt{\log K}$ for $K\gg 1$ \cite{engel1992storage,barkai1992broken,monasson1995weight,mitchison1989bounds}. Such a simplification is not in general possible for weakly differentiable activation functions, as the resulting hidden unit activation distributions will contain a continuous component provided that the activation function is not almost-everywhere constant. 

The fact that small-$K$ output preactivation distributions can have substantial qualitative differences from the $K\to\infty$ Gaussian limit is illustrated by the simple example of ReLU activation functions. If there are only a few hidden units, a substantial fraction of the total probability mass will be concentrated as a Dirac mass at $-\vartheta$, which is a measure-zero event in the infinite-width limit. Such considerations suggest that the behavior of treelike committee machines with only a few hidden units may differ noticeably from that of infinitely-wide networks.

Finally, given a generative model in which the components of the input patterns are independent and identically distributed with mean zero and unit variance, the mean field theory depends on the number of inputs to each hidden unit $N/K$ only through the fact that the distribution of preactivations is taken to be Gaussian. If $N/K$ were finite, this should be a reasonable approximation provided that this ratio is not too small. Alternatively, one could simply take the input distribution to be Gaussian, which can be technically convenient if one aims to provide rigorous proofs \cite{ding2019capacity}. Heuristically, this suggests that our results should carry over to an alternative thermodynamic limit in which one simultaneously takes $N,K\to\infty$ with a large but fixed integer ratio $N/K = \mathcal{O}(1)$.

\section{Gaussian measures, Hermite polynomials, and weak differentiability}\label{app:sec:gauss}

In this appendix, we review relevant background material from the theory of Gaussian measures. Our discussion is a specialization of the more general discussion in Chapter 1 of \citet{bogachev1998gaussian} to the one-dimensional case. We merely seek to summarize the relevant definitions and results, and will not attempt to provide rigorous proofs. 

We let $\gamma$ be the standard Gaussian probability measure on $\mathbb{R}$, which has density $\exp(-x^2/2) / \sqrt{2\pi}$ with respect to Lebesgue measure. We let $\mathcal{L}^{2}(\gamma)$ be the Lebesgue space of functions on $\mathbb{R}$ that are square-integrable with respect to $\gamma$, and, for brevity, denote the norm on this space as $\Vert \cdot \Vert_{\gamma}$. The natural orthonormal basis for $\mathcal{L}^{2}(\gamma)$ is given by the set of Hermite polynomials $\{\He_{k}\}_{k=0}^{\infty}$, which can be defined by the formula
\begin{align}
    \He_{k}(x) = \frac{(-1)^{k}}{\sqrt{k!}} \exp\left(\frac{x^2}{2}\right) \frac{d^{k}}{dx^{k}} \exp\left(-\frac{x^2}{2}\right).
\end{align}
The Hermite polynomials satisfy the recurrence relation
\begin{align}
    \He_{k}'(x) = \sqrt{k} \He_{k-1}(x) = x \He_{k}(x) - \sqrt{k+1} \He_{k+1}(x)
\end{align}
for $k\geq 1$, with $\He_{0} \equiv 1$. For a given function $g \in \mathcal{L}^{2}(\gamma)$, we define its Fourier-Hermite coefficients
\begin{align}
    g_{k} = \int g \He_{k}\, d\gamma, 
\end{align}
and the Fourier-Hermite series
\begin{align}
    g(x) = \sum_{k=0}^{\infty} g_{k} \He_{k}(x),
\end{align}
which is guaranteed to converge in mean-square by the fact that $\Vert g \Vert_{\gamma}^{2} = \sum_{k=0}^{\infty} g_{k}^{2}$ is finite.  

Then, using the recurrence relation $\He_{k}'(x) = \sqrt{k} \He_{k-1}(x)$ and the fact that $\He_{0}' \equiv 0$, we can express the $l^{th}$ weak derivative of $g$ as a formal Fourier-Hermite series
\begin{align}
    g^{(l)}(x) = \sum_{k=l}^{\infty} g_{k} \sqrt{(k)_{l}} \He_{k-l}(x),
\end{align}
where $(k)_{r} = k (k-1) \cdots (k-r+1)$ is the falling factorial. If, for some $r \geq 0$, the sum
\begin{align}
    \Vert g^{(r)} \Vert_{\gamma}^{2} = \sum_{k=r}^{\infty} (k)_{r} g_{k}^{2}
\end{align}
is finite, then the Fourier-Hermite series for $g$ and its weak derivatives up to order $r$ converge in mean-square. The class of functions satisfying this condition is the Sobolev class $\mathcal{H}^{r}(\gamma)$, which has Sobolev norm
\begin{align}
    \Vert g \Vert_{\mathcal{H}^{r}(\gamma)} = \left(\sum_{l=0}^{r} \Vert g^{(l)} \Vert_{\gamma}^{2} \right)^{1/2}. 
\end{align}

Having defined $\mathcal{H}^{r}(\gamma)$ in terms of Fourier-Hermite expansions, we now connect this definition to a more generic notion of weak differentiability. Let $\mathcal{C}^{\infty}_{0}(\mathbb{R})$ be the set of all infinitely-differentiable functions with compact support, and let $p \geq 1$. For a locally integrable function $f$, we define its weak derivative $f'$ as a locally integrable function that satisfies the integration by parts formula
\begin{align}
    \int_{\mathbb{R}} \phi'(x) f(x) \,dx = - \int_{\mathbb{R}} \phi(x) f'(x) \,dx
\end{align}
for every $\phi \in \mathcal{C}^{\infty}_{0}(\mathbb{R})$. The subset of functions in $\mathcal{L}^{2}(\mathbb{R})$ with weak derivatives up to order $r$ of finite $\mathcal{L}^{2}$ norm forms the Sobelev class $\mathcal{H}^{r}(\mathbb{R})$. We can then define the class $\mathcal{H}^{r}_{\mathrm{loc}}(\mathbb{R})$ as the set of all functions $f$ on $\mathbb{R}$ such that $\phi f \in \mathcal{H}^{r}(\mathbb{R})$ for all $\phi \in \mathcal{C}^{\infty}_{0}(\mathbb{R})$. $\mathcal{H}^{r}(\gamma)$ coincides with the class of all functions $f \in \mathcal{H}^{r}_{\mathrm{loc}}(\mathbb{R})$ such that $f$ and its weak derivatives up to order $r$ have finite $\mathcal{L}^{2}(\gamma)$ norm, and the corresponding weak derivatives coincide as well. In one dimension, the criterion that the $(r-1)^{th}$ derivative is differentiable almost everywhere and is equal almost everywhere to the Lebesgue integral of its derivative implies the required weak differentiability condition. Furthermore, by Rademacher's theorem, every function that is locally Lipschitz continuous belongs to $\mathcal{H}^{1}_{\mathrm{loc}}(\mathbb{R})$. 

\section{The Gardner volume of the treelike committee machine}\label{app:sec:gardner}

In this appendix, we give a detailed account of the computation of the Gardner volume of the treelike committee machine using the replica method. As described in the main text, the treelike committee machine \cite{engel1992storage,baldassi2019properties,barkai1992broken,engel2001statistical} is a two-layer neural network with a total of $N$ inputs divided into disjoint groups of $N/K$ among $K$ hidden units, with output
\begin{align}
    y(\mathbf{x}; \{\mathbf{w}_{j}\}, \mathbf{v}, \vartheta) = \sign\big(s(\mathbf{x}; \{\mathbf{w}_{j}\}, \mathbf{v}, \vartheta)\big)
\end{align}
for
\begin{align}
    s(\mathbf{x}; \{\mathbf{w}_{j}\}, \mathbf{v}, \vartheta) &= \frac{1}{\sqrt{K}} \sum_{j=1}^{K} v_{j} g\left(\frac{\mathbf{w}_{j} \cdot \mathbf{x}_{j}}{\sqrt{N/K}}\right) - \vartheta,
\end{align}
where $\mathbf{x}_{j} \in \mathbb{R}^{N/K}$ is the vector of inputs to the $j^{th}$ hidden unit, $\{\mathbf{w}_{j} \in \mathbb{R}^{N/K}\}_{j=1}^{K}$ are the hidden unit weight vectors, $\mathbf{v} \in \mathbb{R}^{K}$ is the fixed readout weight vector, $g$ is the activation function, and $\vartheta \in \mathbb{R}$ is a threshold. We want to characterize the ability of this network to classify a dataset of $P$ independent and identically distributed random examples $\{(\mathbf{x}^{\nu},y^{\nu})\}_{\mu=1}^{P}$, where $\mathbf{x}^{\nu} \in \{-1,1\}^{N}$ and $y^{\nu} \in \{-1,1\}$, in terms of the Gardner volume \cite{gardner1988space,gardner1988optimal}
\begin{align}
    Z_{N,P,K} = \int d\rho(\{\mathbf{w}_{j}\}) \prod_{\nu=1}^{P} \Theta\left( y^{\nu} s(\mathbf{x}^{\nu}; \{\mathbf{w}_{j}\}, \mathbf{v}, \vartheta) - \kappa \right),
\end{align}
where $\rho$ is a measure on the space of hidden unit weights. We will compute the limiting quenched free entropy per weight $f$ in the sequential limit $N,P \to \infty$, $K \to \infty$, with load $N/P \to \alpha \in (0,\infty)$ using the replica trick as
\begin{align}
    f \equiv \lim_{K \to \infty} \lim_{N\to \infty} f_{N,\alpha N,K}
    = \lim_{K \to \infty} \lim_{N \to \infty} \frac{1}{N} \mathbb{E}_{\mathbf{x},y} \log Z_{N,\alpha N,K} 
    = \lim_{n \downarrow 0} \lim_{K \to \infty} \lim_{N \to \infty} \frac{1}{nN} \log \mathbb{E}_{\mathbf{x},y} Z_{N,\alpha N,K}^{n},
\end{align}
where $\mathbb{E}_{\mathbf{x},y}$ denotes expectation over the quenched Bernoulli disorder represented by the dataset. 

We take the elements of $\mathbf{x}^{\nu}$ to be independent and identically distributed, with equal probability of being positive or negative. We allow the distribution of $y^{\nu}$ to be asymmetric, with $\mathbb{P}(y^{\nu} = +1) = 1 - \mathbb{P}(y^{\nu} = - 1) = p$ for some $p \in [0,1]$. We consider the case of spherical weights \cite{gardner1988space,gardner1988optimal,barkai1992broken,engel1992storage,baldassi2019properties}, in which the hidden unit weight vectors are uniformly distributed on the $N/K$-sphere of radius $(N/K)^{1/2}$. The total volume of weight space, which determines the normalizing constant required to make $\rho$ a probability measure, is then $S_{N/K}^{K}$, where
\begin{align}
    S_{D} \equiv \frac{2 \pi^{D/2}}{\Gamma(D/2)} D^{(D-1)/2}
\end{align}
is the surface area of the $D$-dimensional sphere of radius $\sqrt{D}$. We will assume that $\Vert \mathbf{v} \Vert_{2} = \sqrt{K}$, but will not initially impose further conditions on the readout weights or threshold. Finally, we assume that $g \in \mathcal{L}^{2}(\gamma)$. 

Introducing replicas indexed by $a = 1, \ldots, n$, we can write the $n^{th}$ quenched moment of the Gardner volume as
\begin{align}
    \mathbb{E}_{\mathbf{x},y} Z^{n} = \mathbb{E}_{\mathbf{x},y} \int \prod_{a} d\rho(\{\mathbf{w}_{j}^{a} \}) \prod_{a,\nu} \Theta\left( y^{\nu} \left[\frac{1}{\sqrt{K}} \sum_{j} v_{j} g\left(\frac{\mathbf{w}_{j}^{a} \cdot \mathbf{x}_{j}^{\nu}}{\sqrt{N/K}}\right) - \vartheta \right] - \kappa \right).
\end{align}
We observe immediately that the fact that the different patterns are independent and identically distributed implies that
\begin{align}
    \mathbb{E}_{\mathbf{x},y} Z^{n}
    = \int \prod_{a} d\rho(\{\mathbf{w}_{j}^{a} \}) \left[\mathbb{E}_{\mathbf{x},y} \prod_{a} \Theta\left( y \left[\frac{1}{\sqrt{K}} \sum_{j} v_{j} g\left(\frac{\mathbf{w}_{j}^{a} \cdot \mathbf{x}_{j}}{\sqrt{N/K}}\right) - \vartheta\right] - \kappa \right) \right]^{P},
\end{align}
allowing us to simplify our notation by eliminating the pattern index $\nu$. We now consider the local fields
\begin{align}
    h_{j}^{a} \equiv \sqrt{\frac{K}{N}} \mathbf{w}_{j}^{a} \cdot \mathbf{x}_{j}, 
\end{align}
which have mean zero and covariance
\begin{align}
    \cov(h_{j}^{a}, h_{l}^{b}) = \delta_{jl} \frac{K}{N} \mathbf{w}_{j}^{a} \cdot \mathbf{x}_{j}.
\end{align}
In this setting, the natural order parameters are the Edwards-Anderson (EA) order parameters \cite{gardner1988space,talagrand2003spin,mezard1987spin}
\begin{align}
    q_{j}^{ab} \equiv \frac{K}{N} \mathbf{w}_{j}^{a} \cdot \mathbf{w}_{j}^{b} \qquad (a \neq b),
\end{align}
which measure the overlap between the weight vectors of each hidden unit in two different replicas. As we have chosen the weight vectors to lie on the sphere, the self-overlap of each hidden unit is fixed to unity, and the EA order parameters are bounded between negative one and one. In terms of the EA order parameters, we have
\begin{align}
    \cov(h_{j}^{a}, h_{l}^{b}) = \delta_{jl} [\delta_{ab} + q_{j}^{ab} (1-\delta_{ab})].
\end{align}
Then, as each of the local fields is the sum of $N/K$ independent random variables and their covariance is finite, by central limit theorem they converge in distribution as $N \to \infty$ for any fixed $K$ to a multivariate Gaussian with the same mean and covariance \cite{pollard2002user,tong2012multivariate}. We note that this limiting result would alternatively follow by inserting Fourier representations of the delta function to enforce the definition of the variables $h_{j}^{a}$, evaluating the averages over the inputs, and expanding the result to lowest order in $1/N$. 

We then define the function
\begin{align} 
    G_{1}(\{q_{j}^{ab}\}) \equiv \frac{1}{n} \log \mathbb{E}_{y} \mathbb{E}_{h} \prod_{a} \Theta\left(y\left[\frac{1}{\sqrt{K}} \sum_{j} v_{j} g(h_{j}^{a}) - \vartheta\right] - \kappa \right),
\end{align}
where the average $\mathbb{E}_{h}$ is taken over the $q_{j}^{ab}$-dependent Gaussian distribution of the local fields. Introducing Lagrange multipliers $\hat{q}_{j}^{ab}$ to enforce the definitions of the order parameters $q_{j}^{ab}$, we obtain
\begin{align}
    \mathbb{E}_{\mathbf{x},y} Z^{n} = \int \prod_{b<a,j} \frac{dq_{j}^{ab}\, d\hat{q}_{j}^{ab}}{2\pi i K/N} \, \exp\left(-\frac{N}{K} \sum_{b<a,j} q_{j}^{ab} \hat{q}_{j}^{ab} + N n \alpha G_{1}(\{q_{j}^{ab}\}) \right) \int \prod_{a} d\rho(\{\mathbf{w}_{j}^{a} \}) \exp\left(\sum_{b<a,j} \hat{q}_{j}^{ab} \mathbf{w}_{j}^{a} \cdot \mathbf{w}_{j}^{b} \right),
\end{align}
where we have, by convention, rescaled the Lagrange multipliers to absorb the factor of $K/N$ in the definition of the order parameters \cite{gardner1988space}. With the choice that the weight vectors of each branch are uniformly distributed on the $N/K$-sphere of radius $\sqrt{N/K}$, the integral over the weights expands as
\begin{align}
    \frac{1}{S_{N/K}^{K}} \int \prod_{a,j} d\mathbf{w}_{j}^{a} \left[\prod_{a,j} \delta\left(\Vert \mathbf{w}_{j}^{a} \Vert^{2} - \frac{N}{K}\right) \right] \exp\left(\sum_{b<a,j} \hat{q}_{j}^{ab} \mathbf{w}_{j}^{a} \cdot \mathbf{w}_{j}^{b} \right).
\end{align}
To enforce this normalization constraint, we introduce Lagrange multipliers $\hat{E}_{j}^{a}$, allowing us to factor the integrals over the input dimensions of each branch. Furthermore, we note that
\begin{align}
    \lim_{N\to\infty} \frac{K}{N} \log S_{N/K} = \frac{1 + \log(2\pi)}{2}
\end{align}
for any fixed $K$. Then, defining the function
\begin{align} 
    G_{2}(\{q_{j}^{ab}\}, \{\hat{q}_{j}^{ab}\},\{\hat{E}_{j}^{a}\}) &\equiv \frac{1}{2 n K} \sum_{a,j} \hat{E}_{j}^{a} - \frac{1}{n K}\sum_{b<a,j} q_{j}^{ab} \hat{q}_{j}^{ab} - \frac{1+\log(2\pi)}{2} \nonumber\\&\quad + \frac{1}{n K}\sum_{j} \log \int \prod_{a} dw_{j}^{a}\, \exp\left(-\frac{1}{2} \sum_{a} \hat{E}_{j}^{a} (w_{j}^{a})^{2} +  \sum_{b<a} \hat{q}_{j}^{ab} w_{j}^{a} w_{j}^{b} \right),
\end{align}
we can write 
\begin{align}
    \mathbb{E}_{\mathbf{x},y} Z^{n} = \int \prod_{a,j} \frac{d\hat{E}_{j}^{a}}{4\pi i} \int \prod_{b<a,j} \frac{dq_{j}^{ab}\, d\hat{q}_{j}^{ab}}{2\pi i K/N} \exp\left(N n \left[\alpha G_{1}(\{q_{j}^{ab}\}) + G_{2}(\{q_{j}^{ab}\}, \{\hat{q}_{j}^{ab}\},\{\hat{E}_{j}^{a}\})\right]\right)
\end{align}
in the limit $N \to \infty$. In this limit, we can evaluate the integrals over the order parameters and the Lagrange multipliers using the method of steepest descent, which yields an expression for the quenched free entropy as
\begin{align}
    f = \lim_{n \downarrow 0} \extr_{\{q_{j}^{ab}\}, \{\hat{q}_{j}^{ab}\},\{\hat{E}_{j}^{a}\}} \left\{\alpha G_{1}(\{q_{j}^{ab}\}) + G_{2}(\{q_{j}^{ab}\}, \{\hat{q}_{j}^{ab}\},\{\hat{E}_{j}^{a}\})\right\}. 
\end{align}
We note that the function $G_{1}$ represents the energetic contribution to the quenched free entropy, while the function $G_{2}$ represents the entropic contribution. 

\section{Replica-symmetric solution}\label{app:sec:rs}

In this appendix, we study the quenched free entropy derived in Appendix \ref{app:sec:gardner} using a replica-symmetric ansatz \cite{gardner1988optimal,gardner1988space,talagrand2003spin,engel2001statistical,mezard1987spin}. In addition, as we expect the different hidden units to be equivalent to one another after averaging over patterns \cite{engel1992storage,barkai1992broken}, we make the ansatz that the order parameters are the same across all hidden units. Concretely, we make the ansatz
\begin{align}
\begin{cases}
    \hat{E}_{j}^{a} = \hat{E} 
    \\
    q_{j}^{ab} = q
    \\
    \hat{q}_{j}^{ab} = \hat{q}
\end{cases}
\end{align}
which substantially simplifies the saddle point equations. 

\subsection{The replica-symmetric quenched free entropy}

Considering the entropic contribution $G_{2}$, we immediately obtain the simplification
\begin{align}
    G_{2} &= \frac{1}{2} \hat{E} - \frac{1}{2} (n-1) q \hat{q} + \frac{1}{n} \log \int d^{n}w\, \exp\left(-\frac{1}{2} \mathbf{w}^{T} \mathbf{A} \mathbf{w} \right) - \frac{1+\log(2\pi)}{2},
\end{align}
where we have defined the $n \times n$ matrix 
\begin{align}
    \mathbf{A} = (\hat{E}+\hat{q}) \mathbf{I}_{n} - \hat{q} \mathbf{1}_{n}\mathbf{1}_{n}^{T}.
\end{align}
Applying the matrix determinant lemma, we have
\begin{align}
    \det \mathbf{A} = (\hat{E}+\hat{q})^{n} \left(1 - \frac{\hat{q}}{\hat{E}+\hat{q}} n\right),
\end{align}
hence we find that 
\begin{align}
    \lim_{n \downarrow 0} \frac{1}{n} \log \int d^{n}w\, \exp\left(-\frac{1}{2} \mathbf{w}^{T} \mathbf{A} \mathbf{w} \right) = \frac{1}{2} \left[\log \frac{2 \pi}{\hat{E} + \hat{q}} + \frac{\hat{q}}{\hat{E} + \hat{q}} \right].
\end{align}
Then, as we would expect, the entropic contribution to the quenched free entropy is the same as for a perceptron \cite{gardner1988space}, yielding
\begin{align}
    \lim_{n \downarrow 0} G_{2} = \frac{1}{2} \left[ \hat{E} + q \hat{q} + \log \frac{1}{\hat{E} + \hat{q}} + \frac{\hat{q}}{\hat{E} + \hat{q}} - 1\right].
\end{align}
As the Lagrange multipliers $\hat{E}$ and $\hat{q}$ appear only in $G_{2}$, they can easily be eliminated from the saddle point equations, yielding
\begin{align}
    \lim_{n \downarrow 0} G_{2} = \frac{1}{2} \left[\frac{q}{1-q} + \log(1-q) \right]
\end{align}
at the replica-symmetric saddle point. 

We now consider the energetic term $G_{1}$. With the replica- and branch-symmetric ansatz, the covariance matrix of the Gaussian-distributed local fields simplifies to
\begin{align}
    \cov( h_{j}^{a}, h_{l}^{b}) = \delta_{jl} [\delta_{ab} + q (1-\delta_{ab})].
\end{align}
Then, as the local fields $h_{j}$ are independent, the internal fields
\begin{align}
    s^{a} \equiv \frac{1}{\sqrt{K}} \sum_{j} v_{j} g(h_{j}^{a}) - \vartheta
\end{align}
are the sums of $K$ independent random variables, with mean
\begin{align}
    \mu \equiv \mathbb{E}_{h} s^{a} = \bigg[\frac{1}{\sqrt{K}} \sum_{j} v_{j}\bigg] (\mathbb{E} g) - \vartheta
\end{align}
and covariance
\begin{align}
    \cov(s^{a}, s^{b})
    = \frac{1}{K} \sum_{j,l=1}^{K} v_{j} v_{l} \cov(g(h_{j}^{a}), g(h_{l}^{b}))
    = \frac{1}{K} \sum_{j=1}^{K} v_{j}^2 \cov(g(h_{j}^{a}), g(h_{j}^{b}))
    = \cov(g(h^{a}), g(h^{b}))
\end{align}
where we have used the fact the fields are independent and identically distributed across branches and the assumption that $\Vert \mathbf{v} \Vert_{2}^{2} = K$. Then, if $\cov(g(h^{a}), g(h^{b}))$ is finite, which holds for any $g \in \mathcal{L}^{2}(\gamma)$, then the classical central limit theorem implies that the internal fields $s^{a}$ converge in distribution as $K \to \infty$ to a multivariate Gaussian with the mean and variance given above \cite{pollard2002user,tong2012multivariate}. 

Defining the quantity
\begin{align}
    \sigma^2 \equiv \var\left[g(x) \,:\, x \sim \mathcal{N}(0,1)\right] = \Vert g \Vert_{\gamma}^{2} - (\mathbb{E}g)^{2}
\end{align}
and the effective order parameter
\begin{align}
    \tilde{q} = \operatorname{cov}\left[g(x),g(y) \,:\, \begin{bmatrix} x \\ y \end{bmatrix} \sim \mathcal{N}\left(0, \begin{bmatrix} 1 & q \\ q & 1 \end{bmatrix} \right) \right],
\end{align}
we can see from the joint distribution of the local fields $h^{a}$ that we can write the covariance matrix of the internal fields as
\begin{align}
    \cov(s^{a}, s^{b}) = \sigma^2 \delta_{ab} + \tilde{q} (1-\delta_{ab}). 
\end{align}
Then, we can expand the first average in $\exp(n G_{1})$ in terms of the joint characteristic function of $s^{a}$ as
\begin{align}
    \int \prod_{a} \frac{ds^{a}\, d\hat{s}^{a}}{2\pi} \left[\prod_{a} \Theta(-s^{a}-\kappa)\right]  \exp\left(i \sum_{a} s^{a} \hat{s}^{a} - \frac{1}{2} (\sigma^2-\tilde{q}) \sum_{a} (\hat{s}^{a})^{2} - \frac{1}{2} \tilde{q} \bigg[ \sum_{a} \hat{s}^{a} \bigg]^2 + i \mu \sum_{a} \hat{s}^{a} \right).
\end{align}
To evaluate the remaining integrals, we perform a Hubbard-Stratonovich transformation, which is defined via the integral identity \cite{engel2001statistical}
\begin{align}
    \exp\left(-\frac{1}{2} x^2\right) = \int d\gamma(z)\, \exp(-i x z),
\end{align}
to decouple the replicas at the expense of introducing an auxiliary field $z$. From a statistical point of view, we can see that this has the effect of shifting the mean of $s^{a}$ from $\mu$ to $\mu + \sqrt{\tilde{q}} z$, which yields
\begin{align}
    & \int d\gamma(z) \left[  \int \frac{ds\, d\hat{s}}{2\pi} \Theta(-s-\kappa) \exp\left(i s \hat{s} - \frac{1}{2}  (\sigma^2-\tilde{q}) \hat{s}^{2} + i (\mu+\sqrt{\tilde{q}} z) \hat{s} \right) \right]^{n}
    \\
    &\qquad  = \int d\gamma(z) \left[ H\left(\frac{\kappa + \mu + \sqrt{\tilde{q}} z}{\sqrt{\sigma^2 - \tilde{q}}}\right) \right]^{n}, 
\end{align}
where $H(z) = \int_{z}^{\infty} d\gamma(x)$ is the Gaussian tail distribution function. Analogously, we can see that the second term in $\exp(nG_1)$ can be written in a similar form, yielding 
\begin{align}
    \exp(nG_1) = (1-p) \int d\gamma(z) \left[ H\left(\frac{\kappa + \mu + \sqrt{\tilde{q}} z}{\sqrt{\sigma^2 - \tilde{q}}}\right) \right]^{n} + p \int d\gamma(z) \left[ H\left(\frac{\kappa - \mu - \sqrt{\tilde{q}} z}{\sqrt{\sigma^2 - \tilde{q}}}\right) \right]^{n}.
\end{align}
Applying the identity
\begin{align}
    \mathbb{E} \log x = \lim_{n \downarrow 0} \frac{\log(\mathbb{E} x^{n})}{n},
\end{align}
upon passing to the limit $n \downarrow 0$ we obtain the replica-symmetric free entropy
\begin{align}
    f_{\textrm{RS}} = \extr_{q} \bigg\{ & (1-p) \alpha \int d\gamma(z)\, \log H\left(\frac{\kappa + \mu + \sqrt{\tilde{q}} z}{\sqrt{\sigma^2 - \tilde{q}}}\right) + p \alpha \int d\gamma(z)\, \log H\left(\frac{\kappa - \mu - \sqrt{\tilde{q}} z}{\sqrt{\sigma^2 - \tilde{q}}}\right) \nonumber\\&\qquad + \frac{1}{2} \left[\frac{q}{1-q} + \log(1-q)\right] \bigg\}.
\end{align}

\subsection{The replica-symmetric capacity}

The replica-symmetric capacity $\alpha_{\textrm{RS}}$ is determined by the value of $\alpha$ such that $q \uparrow 1$ \cite{gardner1988space,gardner1988optimal,engel1992storage,barkai1992broken,baldassi2019properties,engel2001statistical}. Solving the saddle point equation for $q$ from the replica-symmetric free entropy $f_{\textrm{RS}}$ for $\alpha^{-1}$, we have
\begin{align}
    \frac{1}{\alpha_{\textrm{RS}}} = \lim_{q \uparrow 1} \frac{(1-q)^{2}}{q} \frac{\partial \tilde{q}}{\partial q} \bigg[& (1-p) \int d\gamma(z)\, \frac{1}{H(c_{+})} \frac{1}{\sqrt{2\pi}} \exp\left(-\frac{c_{+}^{2}}{2} \right) \frac{1}{(\sigma^2-\tilde{q})^{3/2}} \left(\kappa + \mu + \frac{z \sigma^2}{\sqrt{\tilde{q}}} \right)
    \nonumber\\
    &\qquad + p \int d\gamma(z)\, \frac{1}{H(c_{-})} \frac{1}{\sqrt{2\pi}} \exp\left(-\frac{c_{-}^{2}}{2} \right) \frac{1}{(\sigma^2-\tilde{q})^{3/2}} \left(\kappa - \mu - \frac{z \sigma^2}{\sqrt{\tilde{q}}} \right) \bigg],
\end{align}
where, for brevity, we write
\begin{align}
    c_{\pm} \equiv \frac{\kappa \pm \mu \pm \sqrt{\tilde{q}} z}{\sqrt{\sigma^2 - \tilde{q}}}.
\end{align}
We can then see that the finiteness of the replica-symmetric critical capacity depends on the analytic properties of $\tilde{q}$ in the limit $q \uparrow 1$. To study the properties of this limit, we make the change of variables $q = 1-\varepsilon$. We generically expect $\tilde{q} \uparrow \sigma^2$, but the way in which $\tilde{q}$ approaches $\sigma^2$ depends on the activation function. As observed by \citet{baldassi2019properties}, for $g(x) = \sign(x)$, $\sigma^2 - \tilde{q} \sim \sqrt{\varepsilon}$, while, for $g(x) = \relu(x)$, $\sigma^2 - \tilde{q} \sim \varepsilon$. As shown in the main text, the asymptotic scaling $\sigma^2 - \tilde{q} \sim \varepsilon$ holds for all $g \in \mathcal{H}^{1}(\gamma)$. We thus make the ansatz
\begin{align}
    \tilde{q} \sim \sigma^2 - \beta \varepsilon^{\ell}
\end{align}
for some parameters $\beta , \ell > 0$. Then, the contribution of the first term in the saddle point equation above to $\alpha_{\textrm{RS}}^{-1}$ is given to leading order as
\begin{align}
    \frac{\varepsilon^{1-\ell/2}}{1-\varepsilon} \frac{\ell (1-p)}{\sqrt{\beta}} \int d\gamma(z)\, \frac{1}{H(c_{+})} \frac{1}{\sqrt{2\pi}} \exp\left(-\frac{c_{+}^{2}}{2} \right) \frac{1}{(\sigma^2-\tilde{q})^{3/2}}  \left(\kappa + \mu + \frac{z \sigma^2}{\sqrt{\sigma^2-\beta\varepsilon^{\ell}}} \right), 
\end{align}
where we have reparameterized $c_{+}$ in terms of $\varepsilon$. In the limit $\varepsilon \downarrow 0$, $c_{+}$ tends to $+\infty$ if $z \geq -(\kappa + \mu) / \sigma$ and to $-\infty$ otherwise. Noting that
\begin{align}
    \frac{1}{H(x)} \sim 
    \begin{cases}
        1 + (2\pi x^{2})^{-1/2} \exp(-x^2/2) \left[1 - x^{-2} + \mathcal{O}(x^{-4}) \right] & x \ll -1 \\
        \sqrt{2 \pi} x \exp(x^2/2) \left[1 + x^{-2} + \mathcal{O}(x^{-4}) \right] & x \gg +1
    \end{cases}, 
\end{align}
we can see that the only non-vanishing contribution comes from the interval $z \geq - (\kappa + \mu) / \sigma$. Thus, to leading order, this term is given as
\begin{align}
    \frac{\varepsilon^{1-\ell}}{1-\varepsilon} \frac{\ell (1-p)}{\beta} \int_{-(\kappa+\mu)/\sigma}^{\infty} d\gamma(z)\, \left(\kappa + \mu + z \sqrt{\sigma^2 - \beta \varepsilon^{\ell}}\right) \left(\kappa + \mu + \frac{z \sigma^2}{\sqrt{\sigma^2-\beta\varepsilon^{\ell}}} \right), 
\end{align}
which we can further approximate as 
\begin{align}
    \ell \varepsilon^{1-\ell} \frac{1-p}{\beta} \int_{-(\kappa+\mu)/\sigma}^{\infty} d\gamma(z)\, \left(\kappa + \mu + \sigma z \right). 
\end{align}
By an identical procedure, we can derive the leading-order contribution to the second term, in which case the non-vanishing contribution to the integral comes from $z \leq (\kappa - \mu) /  \sigma$, yielding
\begin{align}
    \frac{1}{\alpha_{\textrm{RS}}} = \lim_{\varepsilon \downarrow 0} \ell \varepsilon^{1-\ell} \left[\frac{1-p}{\beta} \int_{-(\kappa+\mu)/\sigma}^{\infty} d\gamma(z)\, \left(\kappa + \mu + \sigma z \right) + \frac{p}{\beta} \int_{-\infty}^{(\kappa - \mu) /  \sigma} d\gamma(z)\, \left(\kappa - \mu - \sigma z \right) \right].
\end{align}
We note that this result can alternatively be obtained using the method of \citet{engel1992storage}, which exploits the properties of the function whose extremum with respect to $q$ defines the free entropy $f_{\textrm{RS}}$ to avoid the need to explicitly compute the saddle point equation. 

Thus, we can see that the limit $\varepsilon \downarrow 0$ vanishes for $\ell < 1$, implying divergence of the replica-symmetric capacity. If $\ell \geq 1$, which holds for all functions $g \in \mathcal{H}^{1}(\gamma)$, the capacity remains finite, and we have $\beta = \Vert g' \Vert_{\gamma}^{2}$. We note that the boundary case $\ell=1$, which corresponds to non-zero $\Vert g' \Vert_{\gamma}^{2}$, is special, as the capacity vanishes if if $\ell > 1$. For this class of functions, we therefore obtain
\begin{align}
    \frac{1}{\alpha_{\textrm{RS}}} = \frac{\sigma^2}{\Vert g' \Vert_{\gamma}^{2}} \left[ (1-p) \int_{-(\kappa + \mu) / \sigma}^{\infty} d\gamma(z)\, \left(\frac{\kappa + \mu}{\sigma} + z \right)^2 + p \int_{-(\kappa - \mu) / \sigma}^{\infty} d\gamma(z)\, \left(\frac{\kappa - \mu}{\sigma} + z \right)^2 \right]. 
\end{align}
By inspection, we can see that if $p = 1/2$ and the output distribution is symmetric, $\alpha_{\textrm{RS}}$ is maximized by taking $\mu = 0$. If the condition $\mu = 0$ holds, the formula above simplifies to 
\begin{align}
    \frac{1}{\alpha_{\textrm{RS}}} = \frac{\sigma^2}{\Vert g' \Vert_{\gamma}^{2}} \int_{-\kappa/\sigma}^{\infty} d\gamma(z)\, (\kappa/\sigma + z)^2, 
\end{align}
as given in the main text.

\section{One-step replica-symmetry-breaking solution}\label{app:sec:1rsb}

In this appendix, we consider a one-step replica-symmetry-breaking (1-RSB) ansatz, in which we divide the $n$ replicas into groups of size $m$, known as the Parisi parameter, and allow the overlaps between groups to differ from the overlaps within groups \cite{engel1992storage,engel2001statistical,barkai1992broken,baldassi2019properties,talagrand2003spin,mezard1987spin}. Again, we assume that the order parameters are translation-invariant across branches. We let $q_{0}$ denote the overlaps between replicas in different groups, and $q_{1}$ the overlap between replicas within the same group, with corresponding Lagrange multipliers $\hat{q}_{0}$ and $\hat{q}_{1}$.

\subsection{The 1-RSB quenched free entropy}

With the 1-RSB ansatz, the entropic contribution $G_{2}$ simplifies to
\begin{align}
    G_{2} = \frac{1}{2} \hat{E} - \frac{1}{2} (n-m) q_{0} \hat{q}_{0} - \frac{1}{2} (m-1) q_{1} \hat{q}_{1} + \frac{1}{n} \log \int d^{n}w\, \exp\left(-\frac{1}{2} \mathbf{w}^{T} \mathbf{C} \mathbf{w} \right) - \frac{1 + \log(2\pi)}{2},
\end{align}
where we have defined the $n \times n$ block Toeplitz matrix
\begin{align}
    \mathbf{C} = 
    \begin{pmatrix} 
        \mathbf{A} & \mathbf{B} & \cdots & \mathbf{B} \\ 
        \mathbf{B} & \mathbf{A} & \ddots &  \vdots \\ 
        \vdots & \ddots & \ddots & \\ 
        \mathbf{B} & \cdots & & \mathbf{A}
    \end{pmatrix},
\end{align}
where the $m\times m$ blocks are defined as
\begin{align}
    \mathbf{A} = (\hat{E}+\hat{q}_{1}) \mathbf{I}_{m} - \hat{q}_{1} \mathbf{1}_{m}\mathbf{1}_{m}^{\T}
\end{align}
and
\begin{align}
    \mathbf{B} = -\hat{q}_{0} \mathbf{1}_{m}\mathbf{1}_{m}^{\T},
\end{align}
respectively. Then, as the integral over $\mathbf{w}$ is Gaussian, it can easily be evaluated, yielding
\begin{align}
    \frac{1}{n} \log \int d^{n}w\, \exp\left(-\frac{1}{2} \mathbf{w}^{T} \mathbf{C} \mathbf{w} \right) = \frac{1}{2} \log(2\pi) - \frac{1}{2n} \log \det \mathbf{C}.
\end{align}
To compute the determinant of $\mathbf{C}$, we will use a convenient lemma. For $n/m$ a power of two, we have
\begin{align}
    \det \mathbf{C} = \det(\mathbf{A}-\mathbf{B})^{n/m-1} \det(\mathbf{A}+(n/m-1)\mathbf{B}),
\end{align}
which follows from the identity
\begin{align}
    \det \begin{pmatrix} \mathbf{A} & \mathbf{B} \\ \mathbf{B} & \mathbf{A} \end{pmatrix} = \det(\mathbf{A}-\mathbf{B}) \det(\mathbf{A}+\mathbf{B}),
\end{align}
and induction on $n/m$ in powers of two. By the matrix determinant lemma, we have
\begin{align}
    \det(\mathbf{A}-\mathbf{B}) = (\hat{E}+\hat{q}_{1})^{m} \left(1 + m \frac{\hat{q}_{0}-\hat{q}_{1}}{\hat{E}+\hat{q}_{1}}\right)
\end{align}
and
\begin{align}
    \det(\mathbf{A}+(n/m-1)\mathbf{B}) = (\hat{E}+\hat{q}_{1})^{m} \left(1 + m \frac{(1-n/m)\hat{q}_{0} - \hat{q}_{1}}{\hat{E}+\hat{q}_{1}}\right).
\end{align}
Therefore, for $n/m$ a power of two, we have
\begin{align}
    \frac{1}{n} \log \det \mathbf{C}
    = \log(\hat{E}+\hat{q}_{1}) + \frac{n-m}{nm} \log\left(1 + m \frac{\hat{q}_{0}-\hat{q}_{1}}{\hat{E}+\hat{q}_{1}}\right) + \frac{1}{n} \log\left(1 + \frac{(m-n)\hat{q}_{0} - m\hat{q}_{1}}{\hat{E}+\hat{q}_{1}}\right). 
\end{align}
Assuming the validity of analytic continuation to $n \downarrow 0$, we have
\begin{align}
    \lim_{n \downarrow 0} \frac{1}{n} \log \det \mathbf{C} = \log(\hat{E}+\hat{q}_{1}) - \frac{\hat{q}_{0}}{\hat{E}+\hat{q}_{1} + m (\hat{q}_{0}-\hat{q}_{1})} + \frac{1}{m} \log\left(\frac{\hat{E}+\hat{q}_{1}+m(\hat{q}_{0}-\hat{q}_{1})}{\hat{E}+\hat{q}_{1}}\right).
\end{align}
Therefore, we obtain
\begin{align}
    \lim_{n \downarrow 0} G_{2} &= \frac{1}{2} \hat{E} + \frac{1}{2} q_{1}\hat{q}_{1} + \frac{1}{2} m (q_{0} \hat{q}_{0} - q_{1} \hat{q}_{1}) - \frac{1}{2} \nonumber\\&\qquad + \frac{1}{2} \left[\log\left(\frac{1}{\hat{E}+\hat{q}_{1}}\right) + \frac{\hat{q}_{0}}{\hat{E}+\hat{q}_{1} + m (\hat{q}_{0}-\hat{q}_{1})} + \frac{1}{m} \log\left(\frac{\hat{E}+\hat{q}_{1}}{\hat{E}+\hat{q}_{1}+m(\hat{q}_{0}-\hat{q}_{1})}\right) \right].
\end{align}
We note that this result can alternatively be obtained with substantially more algebra by performing many Hubbard-Stratonovich transformations \cite{engel1992storage}. As the Lagrange multipliers $\hat{E}$, $\hat{q}_{0}$, and $\hat{q}_{1}$ appear only in $G_{2}$, we can eliminate them from the saddle point equations, yielding
\begin{align}
    \lim_{n \downarrow 0} G_{2} = \frac{1}{2} \left[\frac{q_{0}}{1 - q_{1} - m (q_{0}-q_{1})} + \frac{m-1}{m} \log(1-q_{1}) + \frac{1}{m} \log(1 - q_{1} - m (q_{0}-q_{1})) \right]
\end{align}
at the 1-RSB saddle point, which reduces to the replica-symmetric result if we take $q_{0} = q_{1}$.

We now consider the energetic contribution $G_{1}$. As in the replica-symmetric case, the central limit theorem implies that the internal fields
\begin{align}
    s^{a} \equiv \frac{1}{\sqrt{K}} \sum_{j} v_{j} g(h_{j}^{a}) - \vartheta
\end{align}
converge in distribution to a Gaussian as $K \to \infty$. Their mean $\mu$ is the same as before, but now their covariance is given by the block Toeplitz matrix
\begin{align}
    \mathbf{C} = 
    \begin{pmatrix} 
        \mathbf{A} & \mathbf{B} & \cdots & \mathbf{B} \\ 
        \mathbf{B} & \mathbf{A} & \ddots &  \vdots \\ 
        \vdots & \ddots & \ddots & \\ 
        \mathbf{B} & \cdots & & \mathbf{A}
    \end{pmatrix},
\end{align}
with $m\times m$ blocks
\begin{align}
    \mathbf{A} = (\sigma^2-\tilde{q}_{1}) \mathbf{I}_{m} + \tilde{q}_{1} \mathbf{1}_{m}\mathbf{1}_{m}^{\T}
\end{align}
and
\begin{align}
    \mathbf{B} = \tilde{q}_{0} \mathbf{1}_{m}\mathbf{1}_{m}^{\T},
\end{align}
where $\sigma^2 = \mathrm{var}[g(h)]$ as before and the effective order parameter now takes two values 
\begin{align}
    \tilde{q}_{j} = \operatorname{cov}\left[g(x),g(y) \,:\, \begin{bmatrix} x \\ y \end{bmatrix} \sim \mathcal{N}\left(0, \begin{bmatrix} 1 & q_{j} \\ q_{j} & 1 \end{bmatrix} \right) \right], \qquad j = 1,2.
\end{align}
We now want to understand the structure of the joint characteristic function of the replicated internal fields, in terms of which we will expand $G_{1}$, such that we can decouple replicas by performing appropriate Hubbard-Stratonovich transformations. Introducing Lagrange multipliers $\hat{\mathbf{s}}$ and indexing blocks by Greek superscripts, we have
\begin{align}
    \hat{\mathbf{s}} \cdot \mathbf{C} \hat{\mathbf{s}}
    = \sum_{\lambda=1}^{n/m} \hat{\mathbf{s}}^{\lambda} \cdot \mathbf{A} \hat{\mathbf{s}}^{\lambda} + \sum_{\nu \neq \lambda} \hat{\mathbf{s}}^{\nu} \cdot \mathbf{B} \hat{\mathbf{s}}^{\lambda}
    = (\sigma^2-\tilde{q}_{1}) (\hat{\mathbf{s}} \cdot \hat{\mathbf{s}}) + (\tilde{q}_{1} -\tilde{q}_{0}) \sum_{\lambda=1}^{n/m} (\mathbf{1}_{m} \cdot \hat{\mathbf{s}}^{\lambda})^{2} + \tilde{q}_{0} (\mathbf{1}_{n} \cdot \hat{\mathbf{s}})^{2}. 
\end{align}
Then, we can see that we will need to perform one Hubbard-Stratonovich transformation to decouple the $(\mathbf{1}_{n} \cdot \hat{\mathbf{s}})^{2}$ term at the expense of introducing an auxiliary field $z_{0}$, which has the effect of shifting the mean of $s^{a}$ from $\mu$ to $\mu + \sqrt{\tilde{q}_{0}} z_{0}$. To decouple the $(\mathbf{1}_{m} \cdot \hat{\mathbf{s}}^{\lambda})^{2}$ terms, we introduce $n/m$ auxiliary fields $z_{1}^{\lambda}$, which further shifts the mean of $s^{a}$ from $\mu + \sqrt{\tilde{q}_{0}} z_{0}$ to $\mu + \sqrt{\tilde{q}_{0}} z_{0} + \sqrt{\tilde{q}_{1}-\tilde{q}_{0}} z_{1}^{\lambda}$. Then, recognizing that the contribution of each replica within a block to the integral over the corresponding $z_{1}^{\lambda}$ is identical, and that the contribution of each block to the integral over $z_{0}$ is in turn identical, the first average in $\exp(n G_{1})$ is given as
\begin{align}
    & \int d\gamma(z_{0}) \left\{ \int d\gamma(z_{1}) \left[  \int \frac{ds\, d\hat{s}}{2\pi} \Theta(-s-\kappa) \exp\left(i s \hat{s} - \frac{1}{2}  (\sigma^2-\tilde{q}_{1}) \hat{s}^{2} + i (\mu + \sqrt{\tilde{q}_{0}} z_{0} + \sqrt{\tilde{q}_{1}-\tilde{q}_{0}} z_{1}) \hat{s} \right) \right]^{m} \right\}^{n/m}
    \\
    &\qquad  = \int d\gamma(z_{0}) \left\{ \int d\gamma(z_{1}) \left[ H\left(\frac{\kappa + (\mu + \sqrt{\tilde{q}_{0}} z_{0} + \sqrt{\tilde{q}_{1}-\tilde{q}_{0}} z_{1})}{\sqrt{\sigma^2 - \tilde{q}_{1}}}\right) \right]^{m} \right\}^{n/m}.
\end{align}
Analogously, we can see that the second term in $\exp(nG_1)$ can be written in a similar form, yielding 
\begin{align}
    \exp(nG_1) &= (1-p)\int d\gamma(z_{0}) \left\{ \int d\gamma(z_{1}) \left[ H\left(\frac{\kappa + (\mu + \sqrt{\tilde{q}_{0}} z_{0} + \sqrt{\tilde{q}_{1}-\tilde{q}_{0}} z_{1})}{\sqrt{\sigma^2 - \tilde{q}_{1}}}\right) \right]^{m} \right\}^{n/m} \nonumber\\&\qquad + p \int d\gamma(z_{0}) \left\{ \int d\gamma(z_{1}) \left[ H\left(\frac{\kappa - (\mu + \sqrt{\tilde{q}_{0}} z_{0} + \sqrt{\tilde{q}_{1}-\tilde{q}_{0}} z_{1})}{\sqrt{\sigma^2 - \tilde{q}_{1}}}\right) \right]^{m} \right\}^{n/m}.
\end{align}
Therefore, passing to the limit $n \downarrow 0$, we obtain the 1-RSB saddle point free entropy 
\begin{align}
    f_{\textrm{1-RSB}} = \extr_{q_{0},q_{1},m} \bigg\{ & \frac{1}{m} (1-p) \alpha \int d\gamma(z_{0})\, \log \int d\gamma(z_{1}) \left[ H\left(\frac{\kappa + (\mu + \sqrt{\tilde{q}_{0}} z_{0} + \sqrt{\tilde{q}_{1}-\tilde{q}_{0}} z_{1})}{\sqrt{\sigma^2 - \tilde{q}_{1}}}\right) \right]^{m} \nonumber \\&\qquad + \frac{1}{m} p \alpha \int d\gamma(z_{0})\, \log \int d\gamma(z_{1}) \left[ H\left(\frac{\kappa - (\mu + \sqrt{\tilde{q}_{0}} z_{0} + \sqrt{\tilde{q}_{1}-\tilde{q}_{0}} z_{1})}{\sqrt{\sigma^2 - \tilde{q}_{1}}}\right) \right]^{m} \nonumber \\&\qquad + \frac{1}{2} \left[\frac{q_{0}}{1 - q_{1} - m (q_{0}-q_{1})} + \frac{m-1}{m} \log(1-q_{1}) + \frac{1}{m} \log(1 - q_{1} - m (q_{0}-q_{1})) \right] \bigg\}.
\end{align}

\subsection{The 1-RSB capacity}\label{app:subsec:1rsbcapacity}

To determine the capacity under the 1-RSB ansatz, we need to find the value of $\alpha$ such that $q_{1} \uparrow 1$. In this limit, we expect $m \downarrow 0$ such that the non-negative quantity
\begin{align}
    r \equiv \frac{m}{1-q_{1}}
\end{align}
remains finite \cite{engel1992storage,engel2001statistical,barkai1992broken,talagrand2003spin,baldassi2019properties}. We thus re-parameterize the saddle point equations by writing $q_{1} = 1 - \varepsilon$ and $m = r \varepsilon$, which yields
\begin{align}
    f_{\textrm{1-RSB}} = \operatorname*{extr}_{q_{0},\varepsilon,r} \frac{1}{\varepsilon} \bigg\{& \frac{1}{r} (1-p) \alpha \int d\gamma(z_{0})\, \log \int d\gamma(z_{1}) \left[ H\left(\frac{\kappa + (\mu + \sqrt{\tilde{q}_{0}} z_{0} + \sqrt{\tilde{q}_{1}-\tilde{q}_{0}} z_{1})}{\sqrt{\sigma^2 - \tilde{q}_{1}}}\right) \right]^{r\varepsilon} \nonumber \\&\qquad + \frac{1}{r} p \alpha \int d\gamma(z_{0})\, \log \int d\gamma(z_{1}) \left[ H\left(\frac{\kappa - (\mu + \sqrt{\tilde{q}_{0}} z_{0} + \sqrt{\tilde{q}_{1}-\tilde{q}_{0}} z_{1})}{\sqrt{\sigma^2 - \tilde{q}_{1}}}\right) \right]^{r\varepsilon} \nonumber \\&\qquad + \frac{1}{2} \left[\frac{q_{0}}{1 + r (1 - q_{0} -\varepsilon)} + \varepsilon \log(\varepsilon) + \frac{1}{r} \log(1 + r (1 - q_{0} -\varepsilon)) \right] \bigg\},   
\end{align}
where $\tilde{q}_{1}$ is now a function of $\varepsilon$ alone.

To derive a formula for the 1-RSB critical capacity, we follow the method used by \citet{engel1992storage}. This method starts by observing that the quantity inside the curly braces above must vanish in the limit $\varepsilon \downarrow 0$ in order for the extremum with respect to $\varepsilon$ to be well-defined in this limit. This condition gives an implicit expression for $\alpha_{\textrm{1-RSB}}$ as
\begin{align}
    0 = \min_{q_{0},r} \bigg\{\frac{q_{0}}{1 + r (1 - q_{0})} + \frac{1}{r} \log(1 + r (1 - q_{0})) -  \frac{2}{r} \alpha_{\textrm{1-RSB}} \psi(q_{0},r) \bigg\},
\end{align}
where
\begin{align}
    \psi(q_{0},r; \kappa) \equiv -\lim_{\varepsilon \downarrow 0} \bigg\{& (1-p)  \int d\gamma(z_{0})\, \log \int d\gamma(z_{1}) \left[ H(c_{+}) \right]^{r\varepsilon} +  p \int d\gamma(z_{0})\, \log \int d\gamma(z_{1}) \left[ H(c_{-}) \right]^{r\varepsilon} \bigg\}, 
\end{align}
and, for brevity, we write
\begin{align}
    c_{\pm} = \frac{\kappa \pm (\mu + \sqrt{\tilde{q}_{0}} z_{0} + \sqrt{\tilde{q}_{1}-\tilde{q}_{0}} z_{1})}{\sqrt{\sigma^2 - \tilde{q}_{1}}}.
\end{align}
As $\psi \geq 0$ for all $q_{0}$, $r$, and $\kappa$, we can explicitly express the capacity as
\begin{align}
    \alpha_{\textrm{1-RSB}}(\kappa) = \min_{q_{0},r} \left\{ \frac{1}{2\psi(q_{0},r; \kappa)} \left[\frac{r q_{0}}{1 + r (1 - q_{0})} + \log(1 + r (1 - q_{0}))\right] \right\}.
\end{align}
We note that one could obtain the same formula for $\alpha$ as a function of the saddle-point values of $q_{0}$ and $r$ by solving the saddle-point equation for $\varepsilon$ for $\alpha$ and taking the limit $\varepsilon \downarrow 0$. 

We must now evaluate the limit $\varepsilon \downarrow 0$ in the definition of $\psi$. Following our analysis of the RS critical capacity, we focus on the symmetric case $\mu = 0$, and make the ansatz that $\tilde{q}_{0} \sim \sigma^2 - \beta \varepsilon^{\ell}$ for some $\beta,\ell>0$. The assumption of symmetry yields the simplification
\begin{align}
    \psi(q_{0},r;\kappa) = - \lim_{\varepsilon \downarrow 0} \int d\gamma(z_{0})\, \log \int d\gamma(z_{1}) \left[ H\left( \frac{\kappa + \sqrt{\tilde{q}_{0}} z_{0} + \sqrt{\tilde{q}_{1}-\tilde{q}_{0}} z_{1}}{\sqrt{\sigma^2 - \tilde{q}_{1}}} \right) \right]^{r\varepsilon}. 
\end{align}
Expanding the argument of $H$ to leading order in $\varepsilon$, we have
\begin{align}
    \frac{\kappa + \sqrt{\tilde{q}_{0}} z_{0} + \sqrt{\tilde{q}_{1}-\tilde{q}_{0}} z_{1}}{\sqrt{\sigma^2 - \tilde{q}_{1}}} \sim \frac{\kappa + \sqrt{\tilde{q}_{0}} z_{0} + \sqrt{\sigma^2 -\tilde{q}_{0}} z_{1}}{\sqrt{\beta \varepsilon^{\ell}}},
\end{align}
hence the argument of $H$ tends to $+\infty$ for $z_{1} \geq -(\kappa + \sqrt{\tilde{q}_{0}} z_{0})/\sqrt{\sigma^2-\tilde{q}_{0}}$ and to $-\infty$ otherwise. Noting that
\begin{align}
    H(x) \sim 
    \begin{cases}
        1 - (2 \pi x^2)^{-1/2} \exp(-x^2/2) \left[1 - x^{-2} + \mathcal{O}(x^{-4}) \right] & x \ll -1 \\
        (2 \pi x^2)^{-1/2} \exp(-x^2/2) \left[1 - x^{-2} + \mathcal{O}(x^{-4}) \right] & x \gg +1
    \end{cases},
\end{align}
we can then write the argument of the logarithm in the definition of $\psi$ to leading order in $\varepsilon$ as
\begin{align}
    &\int_{-\infty}^{-Q} d\gamma(z_{1})\, \left[ 1 - \frac{1}{\sqrt{2\pi}} \frac{\sqrt{\beta} \varepsilon^{\ell/2}}{\kappa + \sqrt{\tilde{q}_{0}} z_{0} + \sqrt{\sigma^2 -\tilde{q}_{0}} z_{1}} \exp\left(- \frac{(\kappa + \sqrt{\tilde{q}_{0}} z_{0} + \sqrt{\sigma^2 -\tilde{q}_{0}} z_{1})^{2}}{2\beta\varepsilon^{\ell}}\right) \right]^{r\varepsilon} \nonumber\\&\qquad + \int_{-Q}^{\infty} d\gamma(z_{1})\, \left[  \frac{1}{\sqrt{2\pi}} \frac{\sqrt{\beta} \varepsilon^{\ell/2}}{\kappa + \sqrt{\tilde{q}_{0}} z_{0} + \sqrt{\sigma^2 -\tilde{q}_{0}} z_{1}} \exp\left(- \frac{(\kappa + \sqrt{\tilde{q}_{0}} z_{0} + \sqrt{\sigma^2 -\tilde{q}_{0}} z_{1})^{2}}{2\beta\varepsilon^{\ell}}\right) \right]^{r\varepsilon}, 
\end{align}
where we have defined the function
\begin{align}
    Q \equiv \frac{\kappa + \sqrt{\tilde{q}_{0}} z_{0}}{\sqrt{\sigma^2 -\tilde{q}_{0}}}
\end{align}
for brevity. Using the continuity of the logarithm and passing to the limit $\varepsilon \downarrow 0$, this simplifies to 
\begin{align}
    &\int_{-\infty}^{-Q} d\gamma(z_{1})\, + \int_{-Q}^{\infty} d\gamma(z_{1})\,  \lim_{\varepsilon \downarrow 0} \exp\left(- \frac{r(\kappa + \sqrt{\tilde{q}_{0}} z_{0} + \sqrt{\sigma^2 -\tilde{q}_{0}} z_{1})^{2}}{2\beta} \varepsilon^{1-\ell}\right)
\end{align}
for any $\ell > 0$. If $\ell < 1$, the remaining limit tends to unity, hence the argument of the logarithm tends to unity and $\psi$ vanishes, resulting in a divergent 1-RSB capacity. If $g \in \mathcal{H}^{1}(\gamma)$, we have $\ell \geq 1$ and $\beta = \Vert g' \Vert_{\gamma}^{2}$, hence the 1-RSB capacity, like the RS capacity, remains finite. For functions of this class, evaluating the integrals over $z_{1}$, we find that
\begin{align}
    \psi(q_{0},r;\kappa) = - \int d\gamma(z_{0})\, \log\bigg[H(Q)+ R \exp\left(-\frac{1}{2} \frac{r (\kappa + \sqrt{\tilde{q}_{0}} z_{0})^2}{\Vert g' \Vert_{\gamma}^{2} + r (\sigma^2 - \tilde{q}_{0})} \right) H(-Q R) \bigg],
\end{align}
where $Q$ is given as above and we have defined
\begin{align}
    R \equiv \sqrt{\frac{\Vert g' \Vert_{\gamma}^{2}}{\Vert g' \Vert_{\gamma}^{2} + r(\sigma^2-\tilde{q}_{0})}}
\end{align}
for brevity.

To gain some understanding of the behavior of the 1-RSB capacity, we exploit the fact that it is defined as a minimization problem to derive upper bounds by fixing the value of the inter-block overlap $q_{0}$. Trivially, by taking $q_{0} \uparrow 1$ we recover the RS result and the bound $\alpha_{\textrm{1-RSB}} \leq \alpha_{\textrm{RS}}$. If we instead fix $q_{0} = 0$, the problem simplifies dramatically because $\tilde{q}_{0}$ vanishes. Denoting this family of candidate capacities by $\alpha_{\textrm{1-RSB}_{0}}$, we have
\begin{align}
    \alpha_{\textrm{1-RSB}_{0}}(\kappa) = \min_{r \geq 0} \left\{ \frac{\log(1+r)}{2\psi(0,r; \kappa)} \right\},
\end{align}
where
\begin{align}
    \psi(0,r;\kappa) = - \log\left[H\left(\frac{\kappa}{\sigma}\right) + \sqrt{\frac{\Vert g' \Vert_{\gamma}^{2}}{\Vert g' \Vert_{\gamma}^{2} + \sigma^2 r}} \exp\left(-\frac{1}{2} \frac{\kappa^2 r}{\Vert g' \Vert_{\gamma}^{2} + \sigma^2 r} \right) H\left(-\frac{\kappa}{\sigma} \sqrt{\frac{\Vert g' \Vert_{\gamma}^{2}}{\Vert g' \Vert_{\gamma}^{2} + \sigma^2 r}}\right) \right].
\end{align}
Evaluating this expression at $\kappa = 0$ and noting that $\alpha_{\textrm{RS}} = 2 \Vert g' \Vert_{\gamma}^{2} / \sigma^2$, we have
\begin{align}
    \alpha_{\textrm{1-RSB}_{0}} = \min_{s\geq 0} u(s), 
\end{align}
where we have re-expressed the optimization problem in terms of $s \equiv r / \alpha_{\textrm{RS}}$ and defined the function
\begin{align}
    u(s) \equiv \frac{1}{2} \frac{\log(1 + \alpha_{\textrm{RS}} s)}{\log(2) - \log(1+1/\sqrt{1+2s})}.
\end{align}
For all $s > -1/\max\{2,\alpha_{\textrm{RS}}\}$, $u(s)$ is a continuously differentiable transcendental function of $s$, with $u(s=0) = \alpha_{\textrm{RS}}$. For all $0 < \alpha_{\textrm{RS}} \leq 5/2$, $u'(s)$ is positive for all $s > 0$, hence it is minimized at the boundary. For $\alpha_{\textrm{RS}} > 5/2$, $u'(s)$ vanishes for some positive $s$, and the minimum is less than $\alpha_{\textrm{RS}}$. To obtain an asymptotic bound on the 1-RSB capacity for large $\alpha_{\textrm{RS}}$, we can use the fact that
\begin{align}
    \alpha_{\textrm{1-RSB}_{0}} \leq u(1) = \frac{\log(1+\alpha_{\textrm{RS}})}{2\log(3-\sqrt{3})}
\end{align}
for all $\alpha_{\textrm{RS}}$ to obtain the asymptotic
\begin{align}
    \alpha_{\textrm{1-RSB}} = \mathcal{O}(\log\alpha_{\textrm{RS}}).
\end{align}

In summary, the 1-RSB and RS ans{\"a}tze yield the same conditions on the activation function for the capacity to remain finite in the infinite-width limit. For a given activation function, we can in principle determine $\alpha_{\textrm{1-RSB}}$ numerically by solving an explicit two-dimensional minimization problem over $q_{0} \in [0,1]$ and $r_{0} \in [0,\infty)$, though we do not obtain a simple closed-form solution like that for $\alpha_{\textrm{RS}}$. Unlike in the calculation of the RS capacity, we cannot in this case avoid the need to compute the effective order parameter $\tilde{q}_{0}(q_{0})$ for generic values of $q_{0}$. We note that expression for the 1-RSB capacity with $g(x) = \relu(x)$ from \cite{baldassi2019properties} is equivalent in functional form to that presented here. We observe that the first-order conditions on $q_{0}$ and $r$ resulting from this minimization are precisely the saddle-point equations for those order parameters. However, \citet{engel1992storage}'s prescription for expressing the 1-RSB capacity as a minimization problem is advantageous relative to simply solving the saddle point equations as it allows one to derive relatively tractable upper bounds. In particular, the $q_{0} = 0$ family of candidate solutions yields a tighter upper bound on $\alpha_{\textrm{1-RSB}}$ than $\alpha_{\textrm{RS}}$ itself, showing that $\alpha_{\textrm{1-RSB}}$ can grow at most logarithmically with $\alpha_{\textrm{RS}}$.

\section{Computation of the capacity for common activation functions}\label{app:sec:examples}

\begin{figure}[tb]
    \centering
    \includegraphics[width=8.6cm]{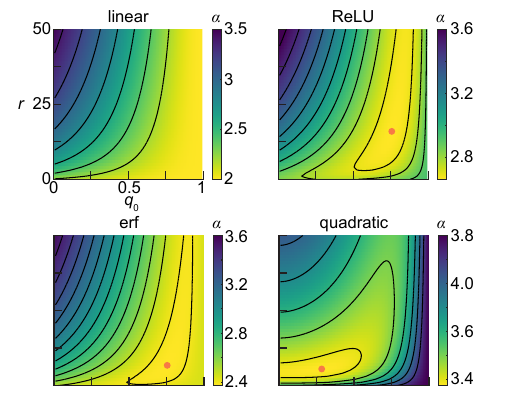}
    \caption{The landscape of the function whose minimum determines the 1-RSB capacity as a function of the inter-block overlap $q_{0}$ and the rescaled Parisi parameter $r \equiv m/(1-q_{1})$ for several example activation functions. In each panel, the value of this function is shown in false color, with the location of minimum indicated by an orange dot.}
    \label{fig:supp:1rsb_minimization}
\end{figure}

In this appendix, we provide details of the computation of the RS and 1-RSB capacities for several commonly-used activation functions. For these examples, we illustrate this minimization problem by plotting its landscape in Figure \ref{fig:supp:1rsb_minimization}. For comparison purposes, we include the landscape for linear activation functions, for which RSB does not occur \cite{talagrand2003spin,gardner1988space,gardner1988optimal,shcherbina2003rigorous}. Numerical computation of the 1-RSB capacity was performed in \textsc{Matlab} 9.6. Integrals with respect to Gaussian measure were estimated using 20-point Gauss-Hermite quadrature \cite{abramowitz1948handbook}, and minimization was performed using the interior-point solver \texttt{fmincon} \cite{byrd2000trust}. These results were then checked against computations performed in \textsc{Mathematica} 12.1 using the numerical integrator \texttt{NIntegrate} and minimizer \texttt{NMinimize} with 100 digits of internal precision. These methods were also used to generate and cross-check the contour plots in Figure \ref{fig:supp:1rsb_minimization}.

\subsection{The rectified linear unit}

The rectified linear unit $\relu(x) \equiv \max\{0,x\}$ is the most commonly-used activation function in modern machine learning appications \cite{lecun2015deep,ramachandran2017searching}, and has weak derivative $\relu'(x) = \Theta(x)$. With our conventions, the Hermite expansion of ReLU is given as
\begin{align}
    \relu(x) = \frac{1}{\sqrt{2\pi}} + \frac{1}{2} \He_{1}(x) + \frac{1}{\sqrt{2\pi}} \sum_{k=1}^{\infty} \frac{(-1)^{k+1}}{2^{k} (2k-1) k!} \sqrt{(2k)!} \He_{2k}(x). 
\end{align}
By direct computation of the required integrals, we have
\begin{align}
    \sigma^2 = \frac{1}{2} - \frac{1}{2\pi} = \frac{\pi-1}{2\pi}
\end{align}
and
\begin{align}
    \Vert \relu' \Vert_{\gamma}^{2} = \Vert \Theta \Vert_{\gamma}^{2} = H(0) = \frac{1}{2}, 
\end{align}
hence we recover the result of \citet{baldassi2019properties} that
\begin{align}
    \alpha_{\textrm{RS}}(\kappa = 0) = \frac{2\pi}{\pi-1} \simeq 2.93388.
\end{align}
For ReLU, we can express $\tilde{q}(q)$ in closed form by direct integration or by summation of the series expansion resulting from the function's Hermite expansion as
\begin{align}
    \tilde{q}(q) = \frac{q}{4} + \frac{q \arcsin(q) + \sqrt{1-q^2} - 1}{2\pi}.
\end{align}
Using this formula, we obtain the estimate
\begin{align}
    \alpha_{\textrm{1-RSB}}(\kappa = 0) \simeq 2.66428
\end{align}
at $(q_{0}^{\ast},r^{\ast}) \simeq (0.75716, 16.63737)$. This result is consistent with the upper bound $\alpha_{\textrm{1-RSB}} \lesssim 2.85021$ resulting from the $q_{0} = 0$ family of candidate capacities.

This estimate of the 1-RSB capacity of a network with ReLU activations is consistent with the $\alpha_{\textrm{1-RSB}} \simeq 2.6643$ reported by \citet{baldassi2019properties} in an update to their Letter. Previous versions of their work reported an erroneous value of $\alpha_{\textrm{1-RSB}} \simeq 2.92$, which does not agree with our estimate of $\alpha_{\textrm{1-RSB}}$ and exceeds the $q_{0}=0$ bound. To estimate the 1-RSB capacity, they solved the saddle-point equations for $q_{0}$ and $r$ numerically rather than directly minimizing over $\alpha_{\textrm{1-RSB}}$. After the appearance of our work in preprint form, they found that incorrect initialization had allowed the solver to converge on the RS saddle point. Their revised estimate is consistent with ours.

\subsection{The Gauss error function}

The Gauss error function
\begin{align}
    \erf(z) = \frac{2}{\sqrt{\pi}} \int_{0}^{z} dx \, \exp(-x^2) = 1 - 2 H(\sqrt{2} z)
\end{align}
is the most analytically convenient of the commonly-used sigmoidal activation functions. It has the Hermite expansion
\begin{align}
    \erf(x) = \frac{2}{\sqrt{3 \pi}} \sum_{k=0}^{\infty} \frac{(-1)^{k}}{3^{k} (2k+1) k!} \sqrt{(2k+1)!} \He_{2k+1}(x), 
\end{align}
which allows us to easily obtain the closed-form expressions
\begin{align}
    \tilde{q}(q) = \frac{4}{3\pi} \sum_{k=0}^{\infty} \frac{(2k+1)!}{3^{2k} (k!)^2 (2k+1)^2} q^{2k+1} = \frac{2}{\pi} \arcsin\left(\frac{2}{3}q\right)
\end{align}
and
\begin{align}
    \sigma^2 = \tilde{q}(1) = \frac{2}{\pi} \arcsin\left(\frac{2}{3}\right). 
\end{align}
Similarly, we can easily work out that
\begin{align}
    \Vert \erf' \Vert_{\gamma}^{2} = \int d\gamma(x)\, \left[\frac{2}{\sqrt{\pi}} \exp(-x^2) \right]^{2} = \frac{4}{\sqrt{5}\pi}.
\end{align}
This yields
\begin{align}
    \alpha_{\textrm{RS}}(\kappa=0) = \frac{4}{\sqrt{5} \arcsin(2/3)} \simeq 2.45140
\end{align}
and
\begin{align}
    \alpha_{\textrm{1-RSB}}(\kappa=0) \simeq 2.37500, 
\end{align}
with $(q_{0}^{\ast},r^{\ast}) \simeq (0.75463, 7.75682)$. This is consistent with our upper bounds for $\alpha_{\textrm{1-RSB}}$, which in this case simply reduce to the RS capacity as it is less than 2.5.

\subsection{The quadratic}

In neuroscientific studies of two-layer network models, expansive activations such as a quadratic are sometimes considered \cite{poirazi2003pyramidal}. With $g(x) = x^2$, we can trivially work out that $\tilde{q}(q) = 2q^2$, hence we have $\sigma^2 = 2$ and $\Vert g' \Vert_{2}^{2} = \Vert 2 x \Vert_{\gamma}^{2} = 4$. Thus, we find that
\begin{align}
    \alpha_{\textrm{RS}}(\kappa=0) = 4
\end{align}
and 
\begin{align}
    \alpha_{\textrm{1-RSB}}(\kappa = 0) \simeq 3.37466, 
\end{align}
with $(q_{0}^{\ast},r^{\ast}) \simeq (0.28452, 6.39299)$. This result is consistent with the $q_{0}=0$ bound of $\alpha_{\textrm{1-RSB}} \lesssim 3.38100$. 

\subsection{The hyperbolic tangent and logistic}

Though it is less analytically convenient than the error function, the hyperbolic tangent and logistic sigmoid $g(x) = (\tanh(x)+1)/2$ are more commonly used in practical applications \cite{lecun2015deep}. We can numerically evaluate the required integrals, yielding $\Vert \tanh \Vert_{\gamma}^{2} \simeq 0.39429$ and $\Vert \tanh' \Vert_{\gamma}^{2} \simeq 0.46440$ and the estimate
\begin{align}
    \alpha_{\textrm{RS}}(\kappa=0) \simeq 2.35561.
\end{align}
As the RS capacity is scale- and shift-invariant, the RS capacity of the logistic function is the same as that of the hyperbolic tangent.

\subsection{The ``Swish" and ``Mish" functions}

Recent experimental works on deep neural networks have proposed various smooth, non-monotonic functions as alternatives to the rectifier unit. One proposal is the product of a logistic function and a linear function, termed ``Swish" \cite{ramachandran2017searching,panigrahi2019effect}:
\begin{align}
    \mathrm{swish}(x;\beta) = \frac{x}{1 + \exp(-\beta x)},
\end{align}
where $\beta$ is a positive parameter. Conventionally, $\beta$ is either fixed to unity or treated as a trainable weight, and yields the limiting behavior $\lim_{\beta \downarrow 0} \mathrm{swish}(x;\beta) = x$, $\lim_{\beta \to \infty} \mathrm{swish}(x;\beta) = \relu(x)$. We can see that $\alpha_{\textrm{RS}}$ is a monotone increasing function of $\beta$, which tends to the perceptron result $\alpha_{\textrm{RS}}(\kappa=0) = 2$ as $\beta \downarrow 0$ and the ReLU result $\alpha_{\textrm{RS}}(\kappa=0) = 2\pi/(\pi-1)$ as $\beta \to \infty$. With $\beta = 1$, we have $\Vert \mathrm{swish}(\cdot\,;1) \Vert_{\gamma}^{2} \simeq 0.31308$ and $\Vert \mathrm{swish}'(\cdot\,;1) \Vert_{\gamma}^{2} \simeq 0.37948$, and the estimate
\begin{align}
    \alpha_{\textrm{RS}}(\kappa=0) \simeq 2.42416.
\end{align}
Another alternative to ReLU is the ``Mish" function, defined as \cite{panigrahi2019effect}
\begin{align}
    \mathrm{mish}(x) = x \tanh \log(1 + \exp(x)), 
\end{align}
for which we have $\Vert \mathrm{mish} \Vert_{\gamma}^{2} \simeq 0.47908$ and $\Vert \mathrm{mish}' \Vert_{\gamma}^{2} \simeq 0.39455$, and the estimate
\begin{align}
    \alpha_{\textrm{RS}}(\kappa=0) \simeq 2.42852.
\end{align}

\subsection{Hermite polynomials}

\begin{figure}
    \centering
    \includegraphics[width=\columnwidth]{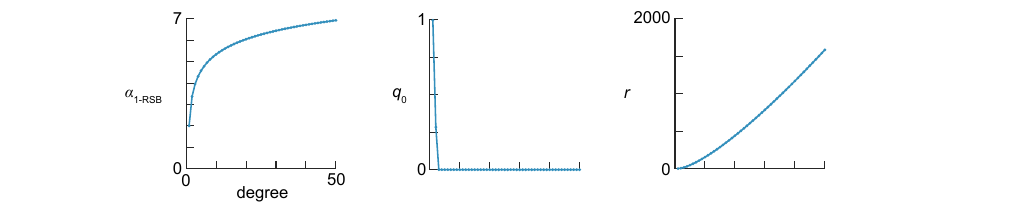}
    \caption{1-RSB solutions for Hermite polynomial activation functions of varying degree. In each panel, the abscissa is the degree of the polynomial. The leftmost panel shows the 1-RSB capacity $\alpha_{\textrm{1-RSB}}$, the middle panel the saddle-point value of the inter-block overlap $q_{0}$, and the rightmost panel the saddle-point value of the rescaled Parisi parameter $r$.}
    \label{fig:supp:hermite_capacity}
\end{figure}

To illustrate how slowly the 1-RSB capacity grows as a function of the RS capacity, we consider Hermite polynomial activations, i.e. 
\begin{align}
    g_{k}(x) = \He_{k}(x)
\end{align}
for $k>0$. For $\He_{k}(x)$, we of course have $\sigma_{k}^2 = 1$, $\Vert \He_{k}' \Vert_{\gamma}^{2} = k$, and $\tilde{q}_{k}(q) = q^{k}$. Thus, the RS capacity at zero margin is simply
\begin{align}
    \alpha_{\textrm{RS}}(k) = 2k.
\end{align}
As we have the simple expression $\tilde{q}(q) = q^{k}$ for any $k$, we can numerically estimate the 1-RSB capacity as a function of degree, yielding the results shown in \ref{fig:supp:hermite_capacity}. Importantly, the 1-RSB capacity grows far slower than linearly with $k$; in particular, it scales roughly as $\log k$ for $k \gg 1$. For this class of activation functions, the saddle point value of the inter-block overlap $q_{0}$ is nearly zero for $k\geq 3$, hence the $q_{0}=0$ upper bound is quite close to the actual estimated value of $\alpha_{\textrm{1-RSB}}$. We observe that the saddle-point value for the rescaled Parisi parameter $r$ scales roughly as a power law for large $k$. 

\section{The replica-symmetric capacity with sparsely active inputs}\label{app:sec:sparse}

In this appendix, we generalize our previous calculation of the replica-symmetric capacity to input distributions with sparsely active input units. Following \citet{gardner1988space}, we consider a distribution with $\mathbb{P}(x_{jk}^{\mu} = +1) = (1+r)/2$, where $r = \mathbb{E} x_{jk}^{\mu} \in [-1,1]$ is the constrained magnetization of the input patterns. We will focus on the limit in which inputs are very sparse ($r \uparrow 1$), but allow the output distribution to potentially be symmetric (that is, we do not assume that $\mathbb{P}(y^{\mu} = +1) = p$ is a function of $r$, as studied by \citet{gardner1988space}). We note that, as shown by \citet{gardner1988space}, it is possible to store more than $\mathcal{O}(N)$ patterns in the limit in which both the input and target output distributions are infinitely sparse. However, such finely tuned matching is not generally realistic in supervised learning tasks. We will follow our previous calculation of the replica-symmetric calculation, noting only where we must make adjustments to account for the non-zero average input magnetization. 

The local fields 
\begin{align}
    h_{j}^{a} \equiv \sqrt{\frac{K}{N}} \mathbf{w}_{j}^{a} \cdot \mathbf{x}_{j}
\end{align}
now have non-zero mean
\begin{align}
    \mathbb{E}_{\mathbf{x}} h_{j}^{a} = r \sqrt{\frac{K}{N}} \sum_{k=1}^{N/K} w_{jk}^{a}
\end{align}
and covariance
\begin{align}
    \cov_{\mathbf{x}}(h_{j}^{a},h_{l}^{b}) = (1-r^2) \delta_{kl} \frac{K}{N} \mathbf{w}_{j}^{a} \cdot \mathbf{w}_{j}^{b}.
\end{align}
We can then see that, in addition to the Edwards-Anderson order parameters $q_{j}^{ab} \equiv (K/N) \mathbf{w}_{j}^{a} \cdot \mathbf{w}_{j}^{b}$, we will need to introduce local magnetizations
\begin{align}
    m_{j}^{a} \equiv \sqrt{\frac{K}{N}} \sum_{k=1}^{N/K} w_{jk}^{a}. 
\end{align}
As noted by \citet{gardner1988space}, the effect of the corresponding Lagrange multiplier on the entropic contribution to the saddle-point free entropy can be neglected in the limit $N\to\infty$, as it is suppressed relative to the contribution from the Lagrange multipliers corresponding to the EA order parameters by a factor of $\sqrt{K/N}$. Thus, the local magnetization affects the free entropy only through its appearance in the energetic term. 

Under a replica- and branch-symmetric ansatz, the defining moments of the local fields are given as
\begin{align}
    \mathbb{E} h_{j}^{a} = r m 
\end{align}
and
\begin{align}
    \cov(h_{j}^{a},h_{l}^{b}) = (1-r^2) \delta_{kl} [\delta_{ab} + q (1-\delta_{ab})]. 
\end{align}
Then, by comparison to our previous results, we can see that we can map all of our reasoning onto the sparse case provided that we take expectations with respect to the modified distribution. In particular, we will have effective order parameters 
\begin{align}
    \tilde{m}(m,r) = \mathbb{E}\Big[ g(x) \,:\, x \sim \mathcal{N}(r m, 1-r^2) \Big], 
\end{align}
\begin{align}
    \sigma^2(m,r) = \var\Big[ g(x) \,:\, x \sim \mathcal{N}(r m, 1-r^2) \Big], 
\end{align}
and
\begin{align}
    \tilde{q}(q,m,r) = \cov\left[g(x),g(y) \,:\, \begin{bmatrix} x \\ y \end{bmatrix} \sim \mathcal{N}\left(r m \begin{bmatrix} 1 \\ 1 \end{bmatrix}, (1-r^2) \begin{bmatrix} 1 & q \\ q & 1 \end{bmatrix} \right) \right]; 
\end{align}
the corresponding quantities in our original calculation are the $r = 0$ special case of these expressions. Defining the average output preactivation
\begin{align}
    \mu \equiv \tilde{m} \frac{1}{\sqrt{K}} \sum_{j=1}^{K} v_{j}  - \vartheta,
\end{align}
the appropriate generalization of the energetic term in the $K\to\infty$ limit from our previous calculation is therefore
\begin{align}
    G_{1} = (1-p) \int d\gamma(z) \, \log H\left(\frac{\kappa + \mu + \sqrt{\tilde{q}}z}{\sqrt{\sigma^2 - \tilde{q}}}\right) + p \int d\gamma(z) \, \log H\left(\frac{\kappa - \mu - \sqrt{\tilde{q}}z}{\sqrt{\sigma^2 - \tilde{q}}}\right),
\end{align}
where in our previous calculation $\mu$ was a constant. 

The replica-symmetric free entropy is then given by
\begin{align}
    f_{\mathrm{RS}} = \extr_{q,m} \Big\{ \alpha G_{1}(q,m) + G_{2}(q) \Big\},
\end{align}
where, as noted above, the entropic term remains unchanged by the introduction of a non-zero magnetization, and is given as
\begin{align}
    G_{2}(q) = \frac{1}{2} \left[\frac{q}{1-q} + \log(1-q)\right].
\end{align}
Under suitable smoothness conditions on the effective order parameters, we can then extract the RS capacity as 
\begin{align}
    \frac{1}{\alpha_{\mathrm{RS}}} = \lim_{q \uparrow 1} \frac{2(1-q)^2}{q} \frac{\partial G_1}{\partial q} \bigg\vert_{m=m^{\ast}},
\end{align}
where $m^{\ast}$ is the solution to the equation
\begin{align}
    \lim_{q \uparrow 1} \frac{\partial G_1}{\partial m} = 0.
\end{align}

Until this point, we have ignored the question of how the local magnetization $m$ should scale with $K$, which affects how $\mu$ scales with $K$. In particular, divergence of $\mu$ corresponds to a trivial committee machine that almost surely predicts the same class for all inputs. In our previous calculation, we handled the scaling of $\mu$ post hoc, as for $r=0$ it is a constant and the subsequent expressions have sensible $|\mu|\to\infty$ limits. Here, however, $\mu$ is a function of the order parameter $m$. As we are interested in the case in which one changes the sparsity of the input distribution while keeping the output distribution fixed, we will demand that $\mu = \mathcal{O}(1)$. We note that solutions with diverging $\mu$ may be optimal if one considers a target distribution that is always either positive or negative, but this is not in general the case. This constraint matches that considered by \citet{gardner1988space} in her analysis of the perceptron, where she demanded that the total magnetization
\begin{align}
    M = \frac{1}{\sqrt{N}} \sum_{k=1}^{N} w_{k} = \frac{1}{\sqrt{K}} \sum_{j=1}^{K} m_{j} = \sqrt{K} m
\end{align}
remain $\mathcal{O}(1)$, corresponding to the scaling $m = \mathcal{O}(K^{-1/2})$. 

As we assume that $\sum_{j=1}^{K} v_{j}^{2} = K$, we should have $|v_{j}| = \mathcal{O}(1)$, hence $\sum_{j=1}^{K} v_{j} = \mathcal{O}(K)$ provided that the readout weights do not exactly sum to zero. We note that this is precisely the case for the classic committee machine readout $v_{j}=1$. As the zero-sum case results in $\mu = -\vartheta$ independent of $m$, we choose to exclude it as it is in this sense trivial. We then define the $\mathcal{O}(1)$ quantity
\begin{align}
    \bar{v} \equiv \frac{1}{K} \sum_{j=1}^{K} v_{j},
\end{align}
in terms of which we have $\mu = \sqrt{K} \bar{v} \tilde{m} - \vartheta$. Following our previous calculation, we choose the threshold such that it cancels the constant component of $\sqrt{K} \bar{v} \tilde{m}$, i.e. we set $\vartheta = \sqrt{K} \bar{v} \tilde{m}_{0}$, where we have defined $\tilde{m}_{0} \equiv \tilde{m}(0,r)$. This prevents $\mu$ from trivially diverging due to the addition of a constant offset to the activations. With these choices, we have $\mu = \sqrt{K} \bar{v} (\tilde{m}-\tilde{m}_{0})$. Therefore, to have $\mu = \mathcal{O}(1)$, we must have $\tilde{m}-\tilde{m}_{0} = \mathcal{O}(K^{-1/2})$.

We can now write down the $K\to\infty$ limit of the saddle point equation for $m$. Defining
\begin{align}
    c_{\pm} \equiv \frac{\kappa \mp \mu \mp \sqrt{\tilde{q}}z}{\sqrt{\sigma^2 - \tilde{q}}}
\end{align}
for brevity, the saddle-point equation for $m$ prior to taking the $q \uparrow 1$ limit is
\begin{align}
    0 &= (1-p) \int d\gamma(z) \, \frac{\sqrt{\sigma^2-\tilde{q}}\phi(c_{-})}{H(c_{-})} \left[ \sqrt{K} \bar{v} \frac{\partial \tilde{m}}{\partial m} - \frac{\kappa + \mu + \sqrt{\tilde{q}} z}{2(\sigma^2-\tilde{q})} \frac{\partial \sigma^2}{\partial m} + \frac{\kappa + \mu + \sigma^2 \tilde{q}^{-1/2} z}{2 (\sigma^2-\tilde{q})} \frac{\partial \tilde{q}}{\partial m}\right]
    \nonumber\\&\quad +
    p \int d\gamma(z) \, \frac{\sqrt{\sigma^2-\tilde{q}} \phi(c_{+})}{H(c_{+})} \left[- \sqrt{K} \bar{v} \frac{\partial \tilde{m}}{\partial m} - \frac{\kappa - \mu - \sqrt{\tilde{q}} z}{2(\sigma^2-\tilde{q})} \frac{\partial \sigma^2}{\partial m} + \frac{\kappa - \mu - \sigma^2 \tilde{q}^{-1/2} z}{2 (\sigma^2-\tilde{q})} \frac{\partial \tilde{q}}{\partial m}\right].
\end{align}
All $K$-dependence in this equation is contained in the $\sqrt{K} \bar{v}$ and $\mu$ terms. Demanding that $\mu = \mathcal{O}(1)$, it simplifies to
\begin{align}
    (1-p) \int d\gamma(z) \, \frac{\sqrt{\sigma^2-\tilde{q}}\phi(c_{-})}{H(c_{-})} \frac{\partial \tilde{m}}{\partial m}
    =
    p \int d\gamma(z) \, \frac{\sqrt{\sigma^2-\tilde{q}} \phi(c_{+})}{H(c_{+})} \frac{\partial \tilde{m}}{\partial m}
\end{align}
in the $K \to \infty$ limit. 

We now have the necessary ingredients to compute the RS capacity in the limit of sparse inputs. We first would like to gain some general understanding of whether the conditions for the RS capacity to remain finite are the same in the sparse limit, assuming that $p$ does not tend to unity with $r$. As in the non-sparse case, this depends on whether the limit 
\begin{align}
    \beta(m,r) = \lim_{q\uparrow 1} \frac{\sigma^2(m,r) - \tilde{q}(q,m,r)}{1-q}
\end{align}
is finite. We assume that $r$ is not exactly equal to one, and that $m$ is bounded. Then, all effective order parameters are the non-sparse ($r=0$) order parameters of a network with a transformed activation function
\begin{align}
    \bar{g}(x) = g(\sqrt{1-r^2} x + r m).
\end{align}
If $g \in \mathcal{L}^{2}(\gamma)$, then $\bar{g}$ must also be in $\mathcal{L}^{2}(\gamma)$, which follows from the fact that $g$ must be exponentially bounded at infinity. The function $\bar{g}$ is weakly differentiable if and only if $g$ is weakly differentiable, which is easy to see based on the fact that the linear transformation $x\mapsto \sqrt{1-r^2} x + r m$ is smooth and invertible. If $g$ is weakly differentiable, then the chain rule for the weak derivative implies that
\begin{align}
    \bar{g}'(x) = \sqrt{1-r^2} g'(\sqrt{1-r^2} x + r m).
\end{align}
Furthermore, if $g'$ has finite $\mathcal{L}^{2}(\gamma)$ norm, than the reasoning applied to $\bar{g}$ above implies that $\bar{g}'$ also has finite norm. Thus, by applying our arguments from the non-sparse case to $\bar{g}$, we find that $\beta$ is finite if the weak derivative of $g$ exists and has finite norm. Therefore, changing the sparsity of the input distribution does not change whether or not the capacity diverges in the infinite-width limit.

We now would like to characterize how the RS capacity of networks with weakly-differentiable activation functions behaves in the sparse limit. However, directly applying the methods that allowed us to evaluate the $q \uparrow 1$ limit in the non-sparse case is more challenging, as we would need to compute the Fourier-Hermite coefficients of the transformed activation $\bar{g}$ for all $r$ and $m$. Instead, we will focus on two simple cases: ReLU and analytic activations with non-vanishing derivative at the origin. These cases cover most activation functions commonly used in neural networks, and provide some insight into how sparsity can affect the capacity.

\subsection{The rectified linear unit}

For $g(x) = \relu(x)$, we can compute the effective local magnetization in closed form, yielding
\begin{align}
    \tilde{m}(m,r) = \sqrt{\frac{1-r^2}{2\pi}} \exp\left(-\frac{(mr)^2}{2(1-r^2)}\right) + \frac{1}{2} mr \left[1 + \erf\left(\frac{mr}{\sqrt{2(1-r^2)}}\right)\right].
\end{align}
We note that the exponential term tends to zero as $r \uparrow 1$ uniformly in $m$, while the second term tends to $m$ as $r \uparrow 1$ if $m>0$, and to zero if $m \leq 0$. 

To obtain an $\mathcal{O}(1)$ value for $\sqrt{K} (\tilde{m}(m,r) - \tilde{m}_{0})$, we adopt the scaling 
\begin{align}
    m = \frac{t}{\sqrt{K}}
\end{align}
for $t = \mathcal{O}(1)$. This yields
\begin{align}
    \lim_{K\to\infty} \sqrt{K} (\tilde{m}(m,r) - \tilde{m}_{0}) = \frac{1}{2} rt
\end{align}
for any $t$ and all $|r| < 1$; other scalings will not yield an $\mathcal{O}(1)$ value. Similarly, we find that
\begin{align}
    \lim_{K\to\infty} \frac{\partial\tilde{m}}{\partial m} = \frac{1}{2} r.
\end{align}

Using the continuity of the effective order parameters in $m$ and the fact that $\relu$ is a positive-homogeneous function, we have
\begin{align}
    \lim_{K\to\infty} \sigma^2(m,r) = \sigma^2(m=0,r) = (1-r^2) \sigma^2(m=0,r=0) = (1-r^2) \frac{\pi-1}{2\pi}
\end{align}
and
\begin{align}
    \lim_{K\to\infty} \tilde{q}(q,m,r) = \tilde{q}(q,m=0,r) = (1-r^2) \tilde{q}(q,m=0,r=0)
\end{align}
for any $|r|<1$. This implies that
\begin{align}
    \beta(m=0,r) = (1-r^2) \beta(m=0,r=0) = \frac{1}{2} (1-r^2).
\end{align}

Then, by the same reasoning we used to take the limit $q\uparrow 1$ to obtain the RS capacity in the non-sparse case, we find that the saddle-point equation for $t$ becomes
\begin{align}
    (1-p)\, u_{1}\left(\frac{\kappa + \bar{v} r t /2}{\sqrt{\sigma^2 (1-r^2)}}\right) = p\, u_{1}\left(\frac{\kappa - \bar{v} r t /2}{\sqrt{\sigma^2 (1-r^2)}}\right),
\end{align}
where by a minor abuse of notation we now write $\sigma^2 = \sigma^2(m=0,r=0)$, and we define
\begin{align}
    u_{n}(x) \equiv \int_{-x}^{\infty} d\gamma(z)\, (x+z)^{n}.
\end{align}
for brevity. Similarly, the re-arranged saddle point equation for $q$ yields
\begin{align}
   \alpha_{\textrm{RS}} = \frac{\pi}{\pi-1} \left[(1-p)\, u_{2}\left(\frac{\kappa + \bar{v} r t /2}{\sqrt{\sigma^2 (1-r^2)}}\right) + p\, u_{2}\left(\frac{\kappa - \bar{v} r t /2}{\sqrt{\sigma^2 (1-r^2)}}\right)\right]^{-1}.
\end{align}

For $p=1/2$, the fact that $u_{1}$ is a monotonically increasing function implies that we must have $t^{\ast}=0$ for any $\kappa$, yielding an RS capacity of
\begin{align}
   \alpha_{\textrm{RS}} = \frac{\pi}{\pi-1} \left[u_{2}\left(\frac{\kappa}{\sqrt{\sigma^2 (1-r^2)}}\right)\right]^{-1}.
\end{align}

\subsection{Analytic activation functions}

We now consider analytic activation functions with non-zero derivative at the origin. As for ReLU, we intuitively expect that the required scaling for $\mu$ to remain $\mathcal{O}(1)$ is $m=t/\sqrt{K}$ for $t = \mathcal{O}(1)$. This intuition may be made more concrete by considering the small-$m$ expansion of $\tilde{m}$ in the limit $r\uparrow 1$:
\begin{align}
    \tilde{m}(m,r)
    &= \int_{-\infty}^{\infty} \frac{dx}{\sqrt{2\pi(1-r^2)}} \exp\left(-\frac{(x-rm)^2}{2(1-r^2)}\right) g(x) 
    \\
    &= \int_{-\infty}^{\infty} \frac{dx}{\sqrt{2\pi(1-r^2)}} \exp\left(-\frac{x^2}{2(1-r^2)}\right) \left[1 + \frac{r x m}{1-r^2} + \mathcal{O}(m^2) \right] g(x) 
    \\
    &= \int_{-\infty}^{\infty} \frac{dx}{\sqrt{2\pi(1-r^2)}} \exp\left(-\frac{x^2}{2(1-r^2)}\right) g(x) \nonumber\\&\qquad +  r m \int_{-\infty}^{\infty} \frac{dx}{\sqrt{2\pi(1-r^2)}} \exp\left(-\frac{x^2}{2(1-r^2)}\right) g'(x) + \mathcal{O}(m^2)
    \\
    &\to g(0) + m g'(0) + \mathcal{O}(m^2),
\end{align}
where we have used the formula for Gaussian integration by parts to obtain the third line, and taken the limit $r \uparrow 1$ on the fourth.

We will proceed under the assumption that $r$ is close enough to unity such that we can expand $\sigma^2(r)$ as a power series in $1-r^2$ by interchanging the expectation with the series expansion of $g(x)$ about the origin. We note that this is justified for sufficiently small $1-r^2$ by the assumption of analyticity. Then, by continuity, we have
\begin{align}
    \lim_{K\to\infty} \sqrt{K} (\tilde{m}(t/\sqrt{K},r) - \tilde{m}_{0}) = g'(0) t + \mathcal{O}(1-r^2),
\end{align}
\begin{align}
    \lim_{K\to\infty} \sigma^2(m,r) = \sigma^2(m=0,r) \equiv \sigma^2(r) = [g'(0)]^{2} (1-r^2) + \mathcal{O}[(1-r^2)^2],
\end{align}
and
\begin{align}
    \lim_{K\to\infty} \tilde{q}(q,m,r) = \tilde{q}(q,m=0,r) \equiv \tilde{q}(q,r) = [g'(0)]^{2} (1-r^2) q + \mathcal{O}[q^2, (1-r^2)^2],
\end{align}
which yields
\begin{align}
    \beta(r) = \lim_{q\uparrow 1} \frac{\sigma^2(r) - \tilde{q}(q,r)}{1-q} = [g'(0)]^{2} (1-r^2) + \mathcal{O}[(1-r^2)^2].
\end{align}
Applying our previous results, we obtain the saddle-point equation for $t$ and the expression for the capacity to leading order in $1-r^2$ as
\begin{align}
    (1-p) \,u_{1}\left(\frac{\kappa + \bar{v} g'(0) t^{\ast}}{\sqrt{[g'(0)]^2 (1-r^2)}} \right) = p \,u_{1}\left(\frac{\kappa - \bar{v} g'(0) t^{\ast}}{\sqrt{[g'(0)]^2 (1-r^2)}}\right)
\end{align}
and
\begin{align}
    \frac{1}{\alpha_{\mathrm{RS}}} = (1-p) u_{2}\left(\frac{\kappa + \bar{v} g'(0) t^{\ast}}{\sqrt{[g'(0)]^2 (1-r^2)}} \right) + p u_{2}\left(\frac{\kappa - \bar{v} g'(0) t^{\ast}}{\sqrt{[g'(0)]^2 (1-r^2)}}\right),
\end{align}
respectively. 

As for ReLU, if $p=1/2$, the saddle-point equation has solution $t^{\ast} = 0$, which yields
\begin{align}
    \alpha_{\textrm{RS}} = \left[u_{2}\left(\frac{\kappa}{\sqrt{[g'(0)]^2 (1-r^2)}} \right) \right]^{-1}.
\end{align}

Comparison of these results reveals an interesting point. With ReLU activation functions, the zero-margin RS capacity remains $2\pi/(\pi-1)$ in the sparse limit. In contrast, the zero-margin RS capacity for an analytic activation function of this type approaches that of the perceptron in this limit. For either, the capacity vanishes at any non-zero margin as $\alpha_{\mathrm{RS}} \sim 1-r^2$, as $u_{2}\sim x^2$ for $x\gg1$. 

The difference in the capacities in the sparse limit for ReLU or analytic activation functions is easy to justify intuitively. For $p=1/2$, one expects the local magnetization to vanish such that the distribution of the output preactivation is symmetric, like that of the target output. Then, the remaining effect of sparsity is the $1-r^2$ variance of the hidden unit preactivations. For ReLU, this simply corresponds to an overall scaling of the output preactivation by $\sqrt{1-r^2}$. For analytic activation functions with non-zero derivative at the origin, we expect terms of quadratic order and higher to be negligible if $1-r^2$ is small enough such that the preactivations are concentrated very near to the origin. This leaves, approximately, a perceptron with an overall scaling factor of $\sqrt{1-r^2} g'(0)$. Then, as the zero-margin capacity is scale-invariant, the zero-margin capacity for ReLU should remain the same as in the non-sparse case, while that for analytic activation functions of this class should approach that of the perceptron. In either case, one expects the capacity at non-zero margins to vanish as the output preactivation concentrates in some $\mathcal{O}(\sqrt{1-r^2})$ neighborhood of zero. Thus, one can obtain the capacities calculated above via heuristic arguments.

\section{Numerical experiments}\label{app:sec:experiments}

The question of how to confront our theory with empirical data raises an important issue in the study of deep networks: the questions of the existence and learnability of solutions to a classification task need not be equivalent \cite{malach2019deeper,goodfellow2016deep,hertz1991introduction}. The Gardner volume seeks to quantify the existence of solutions, agnostic to how the weights might be determined \cite{gardner1988space,gardner1988optimal,barkai1992broken,engel1992storage,engel2001statistical}. For the perceptron, one can prove that the eponymous learning algorithm will find solutions whenever they exist \cite{hertz1991introduction}. However, there do not exist learning algorithms with corresponding convergence guarantees for deep networks \cite{jacot2018neural,belkin2019reconciling,goodfellow2016deep,lecun2015deep}. In the absence of rigorous guarantees, one cannot be sure that a particular learning algorithm will find the solutions which the Gardner volume aims to count. Thus, there exists an important distinction between theories that study the Gardner volume and those that study the storage capacity of networks subject to particular learning rules \cite{knoblauch2010memory,hertz1991introduction,gardner1988optimal,gardner1988space}.

With these considerations in mind, it is important to note that theories of the Gardner volume can be falsified using particular learning algorithms. Concretely, the capacity computed using these methods constitutes a non-rigorous upper bound on the true capacity. Therefore, it is possible to falsify such theories by showing empirically or analytically that a particular learning algorithm can find solutions at loads higher than this predicted bound. However, if one seeks to test the main prediction of our theory---that of diverging or finite capacity in the infinite-width limit---one encounters an important problem: networks with activation functions that are not weakly differentiable are not amenable to optimization via commonly-used gradient-based techniques \cite{lecun2015deep,goodfellow2016deep}. Instead, one must use ad hoc algorithms developed for particular activation functions.

For ``classical" treelike committee machines with sign activation functions and all readout weights equal to unity, the most commonly-used learning algorithms are variants on an algorithm known as least action learning (LAL) \cite{hertz1991introduction,mitchison1989bounds,engel1992storage,engel2001statistical,baldassi2020shaping}. LAL is a greedy heuristic extension of the perceptron learning algorithm: if a training example is classified incorrectly, the perceptron learning rule is applied to the hidden unit with preactivation closest to the threshold among those that ``voted" for the incorrect class. \citet{engel1992storage} found that the empirical capacity of a slight variant of LAL appeared to be around 2 for committee machines of widths 3, 5, and 7, failing to increase with width as predicted by their analysis of the Gardner volume. In particular, the 1-RSB estimate of the capacity with three branches is approximately 3. More recently, \citet{baldassi2020shaping} showed that a version of LAL that operates batchwise can find solutions at loads approaching the 1-RSB estimate, but they only considered the three-branched case. Because the maximum relative change in the output preactivation scales as $1/K$, one expects the speed of learning with LAL to become extremely slow in wide networks. Furthermore, the theoretical capacity with sign activation functions diverges extremely slowly with width, scaling only as $\sqrt{\log{K}}$ \cite{monasson1995weight}.

\begin{figure}
    \centering
    \includegraphics[width=\textwidth]{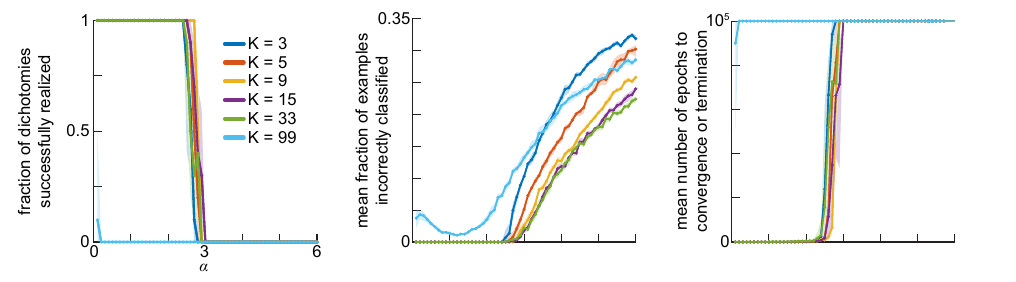}
    \caption{Training treelike committee machines with sign function activations to classify random datasets using batch-LAL. The total number of inputs is $N=990$ throughout, and the abscissa in each panel is the load $\alpha = P/N$. In all panels, solid lines indicate the average over 10 realizations, and shaded patches indicate 95\% confidence intervals of the mean computed via the bias-corrected and accelerated bootstrap method. In the left panel, the ordinate shows the fraction of the 10 realizations at each load for which zero classification error was reached. The center panel shows the mean fraction of examples classified correctly at the end of training at each load. Finally, the right panel shows the mean number of epochs at which training was terminated, either due to early stopping or the fixed threshold.}
    \label{fig:supp:lal}
\end{figure}

We implemented the batch-LAL algorithm proposed in \citet{baldassi2020shaping} in \textsc{Matlab} 9.6 and, following the system size considered in that work, trained a treelike committee machine with sign activation functions with $N=990$ inputs and $K=3$, 5, 9, 15, 33, or 99 hidden units to classify a randomly-generated dataset. We used a learning rate of $\eta = 0.005$ and a batch size of 128 \cite{baldassi2020shaping}, allowing a maximum of 10\textsuperscript{5} epochs with early stopping if zero classification error was reached. As shown in Figure \ref{fig:supp:lal}, the empirical capacities saturate at around 3 rather than increasing with width. This apparent ceiling likely results from the imposed maximum number of training epochs, as we find that the number of epochs required to achieve vanishing error increases superexponentially as the load approaches three from below. Compute time is therefore an important limiting factor in determining the true algorithmic capacity of batch-LAL; these simulations required more than fourteen days of compute time over 32 cores of an HPC node to complete. In short, the batch-LAL algorithm behaves in our hands in much the same way as the variant of LAL proposed by \citet{engel1992storage} did in 1992: if one imposes reasonable constraints the runtime of the algorithm, then one does not observe substantial increases in the empirical capacity with increasing hidden layer width.

\begin{figure}
    \centering
    \includegraphics[width=\columnwidth]{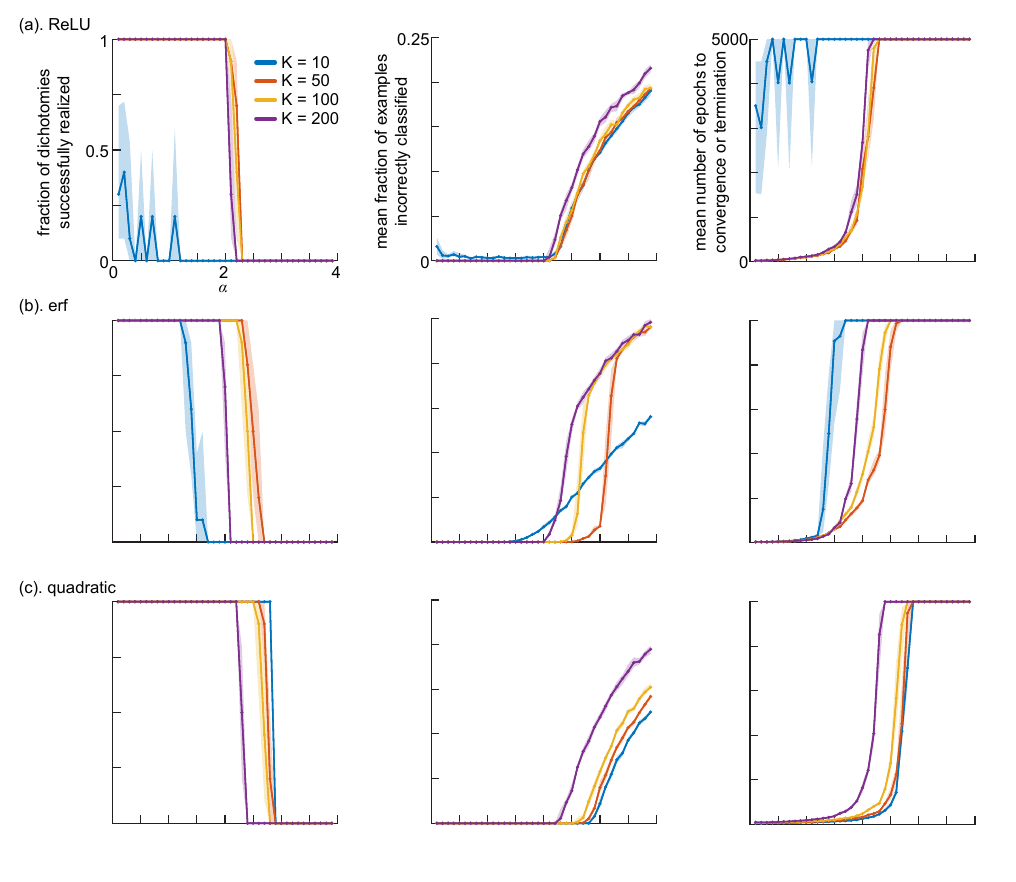}
    \caption{Training treelike committee machines with weakly-differentiable activation functions using stochastic gradient descent on the hinge loss. The total number of inputs is $N=1000$ throughout, and the abscissa in each panel is the load $\alpha = P/N$. In all panels, solid lines indicate the average over 10 realizations, and shaded patches indicate 95\% confidence intervals of the mean computed via the bias-corrected and accelerated bootstrap method. In the left panel, the ordinate shows the fraction of the 10 realizations at each load for which zero classification error was reached. The center panel shows the mean fraction of examples classified correctly at the end of training at each load. Finally, the right panel shows the mean number of epochs at which training was terminated, either due to early stopping or the fixed threshold. Sub-figures (a), (b), and (c) show results for ReLU, erf, and quadratic activation functions, respectively. }
    \label{fig:supp:hinge}
\end{figure}

To study treelike committee machines with weakly differentiable activation functions, we train networks via minimization of the hinge loss. The hinge loss is commonly used for training maximum margin binary classifiers, notably support vector machines \cite{goodfellow2016deep}, and can be optimized using the subgradient methods commonly applied to ReLU networks in contemporary machine learning \cite{lecun2015deep,goodfellow2016deep}. We used this method to train treelike committee machines with $N=1000$ and $K=10$, 50, 100, or 200 with ReLU, erf, or quadratic activation functions. We chose the total number of inputs to be a comparable finite size to that of our LAL simulations while being easily divisible among even numbers of hidden units such that we could set half of the readout weights to $+1$ and the remainder to $-1$, thus satisfying our constraint on the weights and threshold. We implemented the optimization using \textsc{TensorFlow} 2.0 \cite{tensorflow2015} in \textsc{Python} 3.8 using the \textsc{Adam} \cite{kingma2014adam} optimizer with default parameters and a batch size of 32. As shown in Figure \ref{fig:supp:hinge}, we find much the same phenomena in this case as we did for LAL: the empirical capacity appears to be limited chiefly by the maximum number of training epochs allowed. Here, we fixed the maximum number of training epochs to 5,000; each of the 12 simulations reported in Figure \ref{fig:supp:hinge} required between five and seven days of compute time on one NVIDIA Tesla V100 GPU of an HPC node. In all cases, we find empirical capacities that are less than those predicted at 1-RSB, with committee machines with error function activations and 50 hidden units coming the closest to achieving the predicted capacity. Therefore, these experiments fail to falsify our theory.

\end{document}